\documentclass[aps,pra,twocolumn,superscriptaddress]{revtex4-1}
\usepackage{graphicx}
\usepackage{dcolumn}% Align Table columns on decimal point
\usepackage{bm}% bold math
\usepackage{multirow}
\usepackage{ulem}
\usepackage{color}
\usepackage{subfigure}
\usepackage{amsmath}
\usepackage{upgreek}
\usepackage{amsthm,amsmath,amssymb}
\usepackage{lineno}
\usepackage{changes}
\usepackage{array,booktabs}

\begin{document}
%\begin{CJK*}{UTF8}{gbsn}

\preprint{APS/123-QED}

\title{Determination of Land\'{e} $g_J$ factor and Zeeman coefficients in ground-state $^{171}$Yb$^+$ and their applications to quantum frequency standards}

\author{Jize Han}
\affiliation{
State Key Laboratory of Precision Measurement Technology and Instruments, Department of Precision Instrument, Tsinghua University, Beijing 100084, China}%
\affiliation{China Mobile (Suzhou) Software Technology Company Limited, Suzhou 215163, China}

\author{Benquan Lu}
\affiliation{
National Time Service Center, Chinese Academy of Sciences, Xi’an 710600, China}%

\author{Yanmei Yu}
\email{ymyu@aphy.iphy.ac.cn}
\affiliation{Beijing National Laboratory for Condensed Matter Physics, Institute of Physics, Chinese Academy of Sciences, Beijing 100190, China}
\affiliation{University of Chinese Academy of Sciences, Beijing 100049, China}

\author{Jiguang Li}
\email{li\_jiguang@iapcm.ac.cn}
\affiliation{
Institute of Applied Physics and Computational Mathematics, Beijing 100088, China}%

\author{Zhiguo Huang}
\affiliation{China Mobile (Suzhou) Software Technology Company Limited, Suzhou 215163, China}

\author{Jingwei Wen}
\affiliation{China Mobile (Suzhou) Software Technology Company Limited, Suzhou 215163, China}

\author{Ling Qian}
\email{qianling@cmss.chinamobile.com}
\affiliation{China Mobile (Suzhou) Software Technology Company Limited, Suzhou 215163, China}

\author{Lijun Wang}
\email{lwan@mail.tsinghua.edu.cn}
\affiliation{
State Key Laboratory of Precision Measurement Technology and Instruments, Department of Precision Instrument, Tsinghua University, Beijing 100084, China}%
\affiliation{
Department of Physics, Tsinghua University, Beijing 100084, China}%

\date{\today}% It is always \today, today,

\begin{abstract}
We report the determination of the Land\'{e} $g_J$ factor and Zeeman coefficients for the ground-state of $^{171}$Yb$^+$, relevant to microwave quantum frequency standards (QFSs). The $g_J$ factor is obtained by using two independent methods: multiconfiguration Dirac-Hartree-Fock and multireference configuration interaction, yielding a consistent value of 2.002615(70). The first- and second-order Zeeman coefficients are determined as 14,010.78(49) Hz/$\mu$T and 31.0869(22) mHz/$\mu$T$^2$, respectively, based on the calculated $g_J$ factor. These coefficients enable reduced magnetic-field-induced uncertainties, improving the accuracy of the $^{171}$Yb$^+$ microwave QFSs. The results reported in this work also offer potential for improved constraints on variations in fundamental constants through frequency comparisons, and advancing trapped-ion quantum computers based on the ground-state hyperfine splitting of $^{171}$Yb$^+$.
\end{abstract}

\maketitle
%\linenumbers
%\tableofcontents

\section{Introduction}

Microwave quantum frequency standards (QFSs) are among the most widely utilized quantum technologies, offering critical applications in timekeeping and precision measurements. Trapped-ion microwave QFSs represent the next generation of this technology, providing superior stability, precision, and portability \cite{schmittberger2020review}. Among the various trapped-ion systems, \( ^{171} \)Yb\(^+ \) stands out as particularly advantageous compared to other ions such as \( ^{199} \)Hg\(^+ \) \cite{berkeland1998laser, burt2021demonstration} and \( ^{113} \)Cd\(^+ \) \cite{tanaka1996determination, jelenkovic2006high, han2019theoretical, han2019roles, han2021toward, han2022isotope, han2022determination, qin2022high}. This advantage arises from its fiber-friendly cooling and repumping laser wavelengths, which can be generated using compact semiconductor lasers. These unique features make \( ^{171} \)Yb\(^+ \) an attractive candidate for developing compact and practical microwave QFSs \cite{park2007171Yb+, phoonthong2014determination, schwindt2015miniature, schwindt2016highly, mulholland2019compact, mulholland2019laser, han2023cooling, han2024determination}.

In microwave QFSs based on laser-cooled trapped ions, all external fields are minimized except for the magnetic field, which is essential to provide a quantization axis for the ions. This field, commonly referred to as the “C-field” in the QFS community, introduces the most significant frequency shift in trapped-ion microwave QFSs: the second-order Zeeman shift (SOZS). For example, the SOZS contributes over 99\% of the total systematic shifts in laser-cooled Cd$^+$ \cite{miao2021precision, qin2022high} and Yb$^+$ \cite{phoonthong2014determination, warrington2002microwave} microwave QFSs. In laser-cooled Hg$^+$ microwave QFSs, even with only five ions and liquid helium cooling to reduce the required C-field at the cost of stability and portability, the SOZS still accounts for approximately 97\% of the total systematic shifts \cite{berkeland1998laser}. Therefore, a rigorous evaluation of the magnitude and uncertainty of the SOZS is crucial. 

Accurate determination of the SOZS requires precise knowledge of the first- and second-order Zeeman coefficients, $K_Z$ and $K_0$. These coefficients can be determined using extrapolation methods in experiments \cite{hosaka2005optical, rosenband2008frequency, godun2014frequency, brewer2019al+}. Additionally, deriving $K_Z$ and $K_0$ from the $g_J$ factor provides reliable data and serves as a benchmark for extrapolation techniques \cite{hosaka2005optical, rosenband2008frequency, godun2014frequency, brewer2019al+}. Since $g'_I$ \cite{stone2019table} and $A$ \cite{phoonthong2014determination} have been determined with high accuracy, the values of $K_Z$ and $K_0$ primarily depend on the precise determination of the electronic $g_J$ factor. Experimental spectroscopic measurements yield $g_J = 1.998$ \cite{meggers1967second}, while theoretical calculations provide varying predictions: the relativistic coupled-cluster (RCC) method estimates $g_J = 2.002798(113)$ \cite{yu2020ground}, and the time-dependent Hartree-Fock (TDHF) method gives $g_J = 2.003117$ \cite{gossel2013calculation}. Although these theoretical values are in general agreement, they differ by approximately 0.0003. Under typical magnetic field strengths ($3$–$10~\mu$T) \cite{berkeland1998laser, phoonthong2014determination, qin2022high, han2024determination}, this discrepancy in $g_J$ leads to a fractional uncertainty in the SOZS of $(0.8$–$7.9)\times10^{-14}$. Such uncertainty does not meet the accuracy requirements for state-of-the-art $^{171}$Yb$^+$ microwave QFSs \cite{phoonthong2014determination, xin2022laser, han2024determination}. 

Beyond QFS applications, the level structure of $^{171}$Yb$^+$ makes it an excellent candidate for testing the Standard Model and exploring potential new physics. This ion features two optical clock transitions: the E2 transition at 436 nm and the E3 transition at 467 nm. Frequency comparisons between these transitions have established the most stringent constraints to date on the temporal variation of the fine-structure constant $\alpha$, reaching a level of $10^{-19}$ \cite{godun2014frequency, lange2021improved, filzinger2023improved}. Additionally, $^{171}$Yb$^+$ exhibits a 12.6-GHz hyperfine splitting (HFS) that is sensitive to variations in the quark mass to strong-interaction scale ratio, $m_q/\Lambda_{\rm QCD}$ ($^{171}$Yb$^+$: -0.099 vs $^{133}$Cs: 0.002) \cite{flambaum2006dependence, dinh2009sensitivity}. The prospect of conducting frequency comparisons between the two optical clock transitions and the microwave clock transitions of $^{171}$Yb$^+$ within a unified experimental setup is particularly promising. This approach mitigates various common-mode frequency shift uncertainties, including those arising from pressure, gravitational effects, black-body radiation, frequency synchronization, and detection noise, thereby extending the application of the $^{171}$Yb$^+$ microwave QFSs in testing variations of fundamental constants. Moreover, the 12.6-GHz ground-state HFS of $^{171}$Yb$^+$ also encodes as a Qubit level in trapped-ion quantum computers \cite{pino2021demonstration, feng2023continuous, guo2024site, qiao2024tunable}. Thus, improving the accuracy of the Yb$^+$ $g_J$ factor and, consequently, the accuracy of HFS evaluations, could support constraints on the variation of fundamental constants through microwave-optical frequency comparisons and advancements in trapped-ion quantum computers.

In this work, we present the determination of the ground-state Land\'{e} $g_J$ factor and Zeeman coefficients in $^{171}$Yb$^+$ for microwave QFS applications. To achieve an accurate $g_J$ factor, we employ two independent theoretical methods: multiconfiguration Dirac-Hartree-Fock (MCDHF) and multireference configuration interaction (MRCI). These methods yield a consistent value of $g_J = 2.002615(70)$. Based on this result, we derive the first- and second-order Zeeman coefficients as $14,010.78(49)$ Hz/$\mu$T and $31.0869(22)$ mHz/$\mu$T$^2$, respectively. This work improves the accuracy in evaluating the ground-state HFS of $^{171}$Yb$^+$, thereby supporting its applications in microwave QFSs, fundamental constant variation constraints, and trapped-ion quantum computers.

\section{Theory and methods}

In the $6s~^2S_{1/2}~(F=0,m_F=0)\rightarrow (F=1,m_F=0)$ HFS of $^{171}$Yb$^+$, the SOZS in a weak magnetic field regime ($B_0<A/\mu_B$) can be described as \cite{itano2000external},
\begin{equation}
\Delta \nu_{\rm SOZS}=K_0 B_0^2=K_0 (\frac{\Delta \nu_L}{2 K_Z})^2,
\end{equation}
where 
\begin{equation}
K_0=\frac{(g_J-g'_I)^2 \mu_B^2}{2 h^2 A},
\label{eq:K0}
\end{equation}
represents the SOZS coefficient, $g_J$ and $g'_I$ are the electronic and nuclear $g$ factors, $\mu_B$ and $h$ corresponds to the Bohr magneton and the Planck constant, $A$ denotes the ground-state hyperfine constant; and
\begin{equation}
K_Z=\frac{(g_J+g'_I) \mu_B}{2h},
\label{eq:KZ}
\end{equation}
represents the first-order Zeeman coefficient. By incorporating the Larmor frequency difference, \(\Delta \nu_L\), between the Zeeman sublevels \(6s~^2S_{1/2}~(F=0,m_F=0) \rightarrow (F=1,m_F=\pm 1)\), we can calibrate the static magnetic field, \(B_0\), experienced by the ions.

The interaction Hamiltonian between an atom and the magnetic field can be written as \cite{andersson2008hfszeeman}
\begin{eqnarray}
H_m=(\boldsymbol{\rm N}^{(1)}+\boldsymbol{\rm \Delta N}^{(1)})\cdot \boldsymbol{B},
\end{eqnarray}
by choosing the direction of the magnetic field as the $z$-direction and neglecting all diamagnetic contributions in a relativistic frame. The electronic tensor operators of an N-electron atom can be expressed as \cite{cheng1985ab},
\begin{eqnarray}
    \boldsymbol{\rm N}^{(1)}&=&\sum^N_{i=1}\boldsymbol{\rm n}^{(1)}(i)=\sum^N_{i=1}-i\frac{\sqrt{2}}{2\alpha}r_i(\boldsymbol{\alpha}_i\boldsymbol{C}^{(1)}(i))^{(1)},\nonumber\\
    \boldsymbol{\rm \Delta N}^{(1)}&=&\sum^N_{i=1}\boldsymbol{\rm \Delta n}^{(1)}(i)=\sum^N_{i=1}\frac{g_s-2}{2}\beta_i\boldsymbol{\Sigma}_i,
\end{eqnarray}
where $\boldsymbol{\Sigma}_i$ represents the relativistic spin matrix, $i$ denotes the imaginary unit, and the term $\boldsymbol{\rm \Delta N}^{(1)}$ corresponds to the Schwinger QED correction. The $g_s=2.00232$ is the electron spin $g_J$ factor, which is adjusted by a factor of 1.001160 to account for quantum electrodynamics (QED) effects. The uncertainty in $g_s$ is significantly smaller than the current many-body calculation accuracy and therefore can be negligible. The interaction Hamiltonian $H_m$ can be treated in first-order perturbation theory in a weak magnetic field situation. The $g_J$ factor is defined as
\begin{equation}
    g_J=2\frac{\langle \Gamma P M_J J || \boldsymbol{\rm N}^{(1)}+\boldsymbol{\rm \Delta N}^{(1)} || \Gamma P M_J J \rangle}{\sqrt{J(J+1)}},
\end{equation}
where $P$ is the parity, $J$ is total angular momentum, $M_J$ is the component along the $z$ direction of $J$, $\Gamma$ represents another appropriate angular momentum, and $| \Gamma J P M_J \rangle$ is the atomic state function (ASF) which can be obtained using the MCDHF and MRCI methods.

\subsection{The MCDHF method}

\begin{figure}
\centering
\resizebox{0.45\textwidth}{!}{
\includegraphics{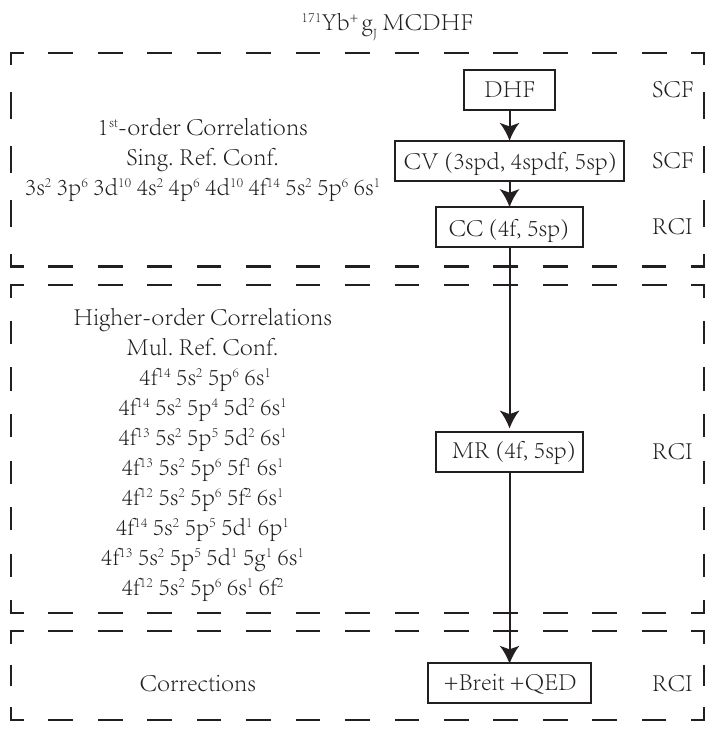}
}
\caption{The computational model employed in the MCDHF method incorporates both first-order and higher-order electron correlations. The first-order electron correlations encompass the CV and CC correlations. The higher-order electron correlations are considered by the MR-SD method. Corrections for the Breit interaction and QED effects are included in the final model.}
\label{fig:mcdhf}
\end{figure}

The MCDHF method \cite{jonsson2022introduction} is implemented in the {\sc Grasp} package \cite{jonsson2023grasp}. The ASF in the MCDHF method are approximate eigenfunctions of the Dirac-Coulomb-Breit Hamiltonian describing an atomic system,
\begin{eqnarray}
H_{\rm DCB}&=&\sum_i [c~\bm{\alpha}_i \cdot \bm{p}_i+(\beta_i-1) c^2+V_{\rm nuc}(r_i) ] +\sum_{i<j}^N\frac{1}{r_{ij}}\nonumber\\
&&-\frac{1}{2r_{ij}}[\bm{\alpha}_i \cdot \bm{\alpha}_j +\frac{( \bm{\alpha}_i\cdot \bm{r}_{ij})(\bm{\alpha}_j\cdot \bm{r}_{ij})}{r_{ij}^2}],
\end{eqnarray}
where $ \bm \alpha$ and $\beta$ represent the Dirac matrices, $\bm p_i$ is the momentum operator, $r_{ij}$ is the distance between electrons $i$ and $j$, and $V_{\rm nuc}(r)$ is the nuclear potential results from a nuclear charge density given by a two-parameter Fermi distribution function \cite{parpia1992relativistic}. The last term is the Breit interaction in the low-frequency approximation. The ASF is a linear combination of configuration state functions (CSFs) \cite{grant2007relativistic},
\begin{eqnarray}
| \Gamma P J M_J \rangle=\sum^{N_{\rm CSF}}_{i=1}c_i|\Gamma_i P J M_J\rangle.
\end{eqnarray}
where $c_i$ are the expansion (mixing) coefficients. The CSFs are the linear combinations of one-electron Dirac orbital products.

The active set approach is employed to generate CSFs, with the selection of CSFs determining the electron correlation effects to be considered. The computational model is outlined in Fig. \ref{fig:mcdhf}. The calculations begin with the Dirac-Hartree-Fock (DHF) approximation, optimizing the occupied spectroscopic orbitals for the reference configuration $\{[\rm Ne]3s^23p^63d^{10}4s^24p^64d^{10}4f^{14}5s^25p^6~6s\}$. In this configuration, the outermost $6s$ orbital is treated as valence electron, while the remaining orbitals are considered core electrons. Valence-core ($n \geq 3$) correlations (CV) are included in the self-consistent field (SCF) calculations, labeled as “CV” in Table \ref{tab:mcdhf}. These correlations are accounted for using single and restricted double (SrD) substitutions from the $n \geq 3$ core orbitals, where only one electron in the core orbital is allowed to be excited to the correlation orbitals. The correlation orbitals are systematically expanded to $\{14s, 13p, 12d, 12f, 10g, 10h, 8i\}$, with only the latest added orbitals optimized in each step. Core-core (CC) correlations are considered in the relativistic configuration interaction (RCI) calculations, where only the mixing coefficients are varied. The CC correlations are incorporated using CSFs generated via single and double (SD) substitutions from the $4f^{14}5s^25p^66s$ core orbitals to the largest correlation orbitals, as indicated by `CC' in Table \ref{tab:mcdhf}. 

\begin{table*}
\caption{The MCDHF calculations of $g_J$ factors using different levels of modeling. NCSF denotes the number of configuration state functions for each model.
\label{tab:mcdhf}} 
{\setlength{\tabcolsep}{8pt}
\begin{tabular}{lllllllll}\hline\hline\addlinespace[0.1cm]
Model	&	Reference configurations&	Correlation orbitals	&	NCSF	&	$g_J$	\\\hline
DHF 	&	$\{5s^25p^66s\}$	&\{5s,5p,6s\}	&	1	&	2.002240 	\\[+1ex]
CV	    &	$\{3s^23p^63d^{10}4s^24p^64d^{10}4f^{14}5s^25p^66s\}$	&	\{14s,13p,12d,12f,10g,10h,8i\}	&	25618	&	2.002884 	\\[+1ex]
CC	&	$\{4f^{14}5s^25p^66s\}$	&	\{14s,13p,12d,12f,10g,10h,8i\}	&	170109	&	2.002528 	\\[+1ex]
MR-I	&	$\cup \{4f^{14}5s^25p^45d^26s;~4f^{13}5s^25p^55d^26s\}$	&	\{8s,7p,7d,6f,6g,6h\}	&	851089	&	2.002578 	\\[+1ex]
MR-II	&	$\cup \{4f^{13}5s^25p^65f^16s;~~4f^{12}5s^25p^65f^26s;$	&	\{8s,8p,7d,8f,6g,6h\}	&	2654356	&	2.002635 	\\
	&	$~~~4f^{14}5s^25p^55d^16p;~4f^{13}5s^25p^55d^15g^16s;$	&		&		&		\\
	&	$~~~4f^{12}5s^25p^66s6f^2\}$	&		&		&		\\[+1ex]
Final  &  & & & 2.002626 \\
Uncertainty  &  & & & 0.000057	 \\
\hline\hline
	\end{tabular}}
\end{table*}

Higher-order electron correlations in the MCDHF calculations are efficiently incorporated through SD excitations from multi-reference (MR) configurations \cite{li2012effects, zhang2017theoretical}. In this approach, CSFs with significant mixing coefficients from the first-order correlations are selected to construct the MR configuration set. SD excitations from this MR configuration set capture the dominant higher-order electron correlation effects. Initially, configurations $\{4f^{14}5s^25p^45d^26s;$ $4f^{13}5s^25p^55d^26s\}$, with mixing coefficients greater than 0.03, are included and labeled as `MR-I'. Subsequently, configurations $\{4f^{13}5s^25p^65f^16s;$ $4f^{12}5s^25p^65f^26s;$ $4f^{14}5s^25p^55d^16p;$ $4f^{13}5s^25p^55d^15g^16s;$ $4f^{12}5s^25p^66s6f^2\}$, with mixing coefficients between 0.025 and 0.03, are added and referred to as `MR-II'. In the MR model, only two layers of correlation orbitals are utilized to generate CSFs due to their rapid convergence. These configurations are incorporated into the `CC' calculation for the reference configuration, as summarized in Table \ref{tab:mcdhf}. Finally, Breit interaction and QED effects (vacuum polarization and self-energy) are included within the RCI procedure in the MR-II model. 

The $g_J$ factors obtained from different computational models based on the MCDHF method are presented in Table \ref{tab:mcdhf}. The CV correlation is the dominant effect, while the CC correlation and higher-order correlations exhibit a canceling effect, consistent with previous observations in the HFS \cite{zhang2017theoretical}. Incorporating the primary CV, CC, and higher-order correlations yields a $g_J$ of 2.002635 from the MCDHF calculations. Breit interaction and QED effects contribute corrections of approximately $-0.000010$ and $0.000001$, respectively, refining the $g_J$ to 2.002626. 

A comprehensive evaluation of the upper and lower bounds of the $g_J$ using the MCDHF method is conducted to quantify the calculation uncertainty. The calculated $g_J$ exhibits an increasing trend with the MR configurations expansion. The upper bound of the $g_J$, constrained by the convergence of the MR expansion, is estimated to be less than the difference between the $g_J$ values obtained from the MR-II and MR-I configurations, which is 0.000057. Electron correlations from the $n=1, 2, 3$ shells and the $4s$, $4p$, and $4d$ orbitals, could reduce the $g_J$. Contributions of the CC and higher-order correlations from the $4f$, $5s$, and $5p$ orbitals are calculated as $-0.000356$ and $0.000107$, respectively, as shown in Table \ref{fig:mcdhf}, indicating a cancellation effect of at least 30\%. The CC contributions from the $4s$, $4p$, and $4d$ orbitals are calculated to be $-0.000076$. Considering the cancellation effect, the lower bound of the $g_J$ due to CC and higher-order electron correlations among electrons with $n\le 4d$ is estimated to be less than $-0.000057$. Other sources of uncertainty are within this range. Consequently, the $g_J$ derived from the MCDHF calculations is determined to be $2.002626(57)$, where the value in parentheses represents the estimated uncertainty.

\subsection{The MRCI method}

\begin{table*}
\caption{The general active space model in the MRCI calculations. \label{tab:GAS}} 
	{\setlength{\tabcolsep}{12pt}
		\begin{tabular}{lccccc}\hline\hline\addlinespace[0.2cm]
 Model	&	Min.	&	Max.	&	\# of Kramer pair 	&	Function type	&	Excitation rank	\\ \hline \addlinespace[0.1cm]
e23-SD	&	7	&	8	&	4	&	$5s^25p^6$, core	&	S	\\
&	21	&	22	&	7	&	$4f^{14}$, core	&	S	\\
&	21	&	23	&	1	&	$6s$,valence	&	SD	\\
&	23	&	23	&	m	&	virtual	&		\\ [+1ex]
e33-SD	&	9	&	10	&	5	&	$4d$, core	&	S	\\
&	17	&	18	&	4	&	$5s^25p^6$, core	&	S	\\
&	31	&	32	&	7	&	$4f^{14}$, core	&	S	\\
&	31	&	33	&	1	&	$6s$,valence	&	SD	\\
&	33	&	33	&	m	&	virtual	&		\\ [+1ex]
e23-SDT-I	&	7	&	8	&	4	&	$5s^25p^6$, core	&	S	\\
&	20	&	22	&	7	&	$4f^{14}$, core	&	SD	\\
&	20	&	23	&	1	&	$6s$,valence	&	SDT	\\
&	23	&	23	&	m	&	virtual	&		\\ [+1ex]
e23-SDT-II	&	6	&	8	&	4	&	$5s^25p^6$, core	&	SD	\\
&	20	&	22	&	7	&	$4f^{14}$, core	&	SD	\\
&	20	&	23	&	1	&	$6s$,valence	&	SDT	\\
&	23	&	23	&	m	&	virtual	&		\\ \hline\hline		
	\end{tabular}}
\end{table*}

The MRCI method is implemented within the KR-CI module of the {\sc Dirac} package \cite{knecht2008large, jensen_h_j_aa_2022_6010450}. The computational procedure begins with a closed-shell DHF calculation of Yb$^{2+}$ using the Dirac-Coulomb-Gaunt Hamiltonian. This Hamiltonian is expressed as
\begin{eqnarray}
H_{\rm DCG}&=&\sum_i [c~\bm{\alpha}_i\cdot \bm{p}_i+(\beta_i-1) c^2+V_{\rm nuc}(r_i) ] + \nonumber \\
&& \sum_{i<j}\bigg[\frac{1}{r_{ij}}-\frac{ {\bm \alpha}_i \cdot {\bm \alpha}_j }{2r_{ij}}\bigg]. \label{DCG}
\end{eqnarray}
The last term in Eq. (\ref{DCG}) corresponds to the Gaunt interaction, which constitutes the primary component of the Breit interaction. The Gaunt interaction is only included in the DHF calculation. We have employed Dyall's core-valence triple-zeta and quadruple-zeta basis sets, which have 30$s$, 24$p$,16$d$, 11$f$, 3$g$, 2$h$ functions (dyall.cv3z) and 35$s$, 30$p$, 19$d$, 13$f$, 5$g$, 4$h$, and 2$i$ functions (dyall.cv4z) \cite{gomes2010relativistic}. We have also used the doubly diffusion augmented basis set of dyall.cv4z, as denoted by `d-aug-dyall.cv4z'. Besides, we have carried out the densification in the $s$, $p$, $d$, $f$, and $g$ spaces of the dyall.cv4z basis by using the technique way suggested in Ref. \cite{hubert2022electric}, which increase the basis size up to 69$s$, 59$p$, 37$d$, 25$f$, 9$g$, 4$h$, and 2$i$ functions. 

The MRCI calculation is performed after the DHF calculation. The MRCI method is based on the general active space (GAS) concept \cite{fleig2001generalized, fleig2003generalized, fleig2006generalized}. Through defining the minimum (Min.) and maximum (Max.) numbers of the accumulated electron occupation, the number (\#) of the Kramer pair, and the allowed excitation ranks that can be single (S), the single and double (SD), and the single, double, and triple (SDT), we choose four different CI wave-function models that are illustrated in Table \ref{tab:GAS}. Firstly, we take the outmost 23 electrons into account, which includes the $5s^25p^64f^{14}$ core and the $6s$ valence electron, as labeled by `e23-SD'. Then we add the $4d$ core, which leads to a model with the outmost 33 electrons, as labeled by `e33-SD'. Next, the allowed excitation ranks are increased to bring the triple electronic correlations, which are labeled by `e23-SDT-I(II)'. The remaining unoccupied spinors consist of the virtual subspace. The high-lying virtual orbitals with energy larger than $m$ a.u. as denoted by `$< m$', are truncated, which is tested to ensure the convergence of the results with such operation.

\begin{table*}
	\caption{The MRCI calculations of $g_J$ performed at different levels of modeling.} \label{tab:mrci}
	{\setlength{\tabcolsep}{8pt}
		\begin{tabular}{llllll}\hline\hline\addlinespace[0.2cm]
Index	&	Model	&	Basis	&	NCSF	&	active virtual spinors 	&	$g_J$	\\ \hline \addlinespace[0.2cm]
(1)	&	e23-SD $<$ 5	&	dyall.cv3z	&	34163	&	$\{10s,10p,8d,7f,5g,6h\}$	&	2.002857 	\\ 
(2)	&	e23-SD $<$ 5 $^{\ast}$	&	dyall.cv3z	&	34163	&	$\{10s,10p,8d,7f,5g,6h\}$	&	2.002850 	\\ 
(3)	&	e33-SD $<$ 5	&	dyall.cv3z	&	50155	&	$\{10s,10p,8d,7f,5g,6h\}$	&	2.002868 	\\ \addlinespace[0.2cm]
(4)	&	e23-SD $<$ 10	&	dyall.cv4z	&	96626	&	$\{12s,12p,9d,9f,7g,7h,7i\}$	&	2.002863 	\\ 
(5)	&	e23-SD $<$ 15	&	dyall.cv4z	&	113980	&	$\{13s,13p,10d,9f,7g,7h,7i\}$	&	2.002855 	\\ 
(6)	&	e23-SD $<$ 50	&	dyall.cv4z	&	210214	&	$\{14s,14p,11d,11f,8g,8h,8i\}$	&	2.002862 	\\ \addlinespace[0.2cm]
(7)	&	e23-SD $<$ 10	&	d-aug-dyall.cv4z	&	232734	&	$\{14s,14p,11d,11f,9g,9h,9i\}$	&	2.002852 	\\ 
(8)	&	e23-SD $<$ 10	&	densified-dyall.cv4z	&	235168	&	$\{14s,14p,11d,11f,9g,9h,8i\}$	&	2.002855 	\\ \addlinespace[0.2cm]
(9)	&	e23-SDT-I $<$10	&	dyall.cv4z	&	37446858	&	$\{12s,12p,9d,9f,7g,7h,7i\}$	&	2.002604 	\\ 
(10)	&	e23-SDT-I $<$10 $^{\ast}$	&	dyall.cv4z	&	37446858	&	$\{12s,12p,9d,9f,7g,7h,7i\}$	&	2.002636 	\\ 
(11)	&	e23-SDT-II $<$10	&	dyall.cv4z	&	86600398	&	$\{12s,12p,9d,9f,7g,7h,7i\}$	&	2.002585 	\\ 
(12)	&	e23-SDT-I $<$10	&	d-aug-dyall.cv4z	&	262470976	&	$\{14s,14p,11d,11f,9g,9h,9i\}$	&	2.002659 	\\ \addlinespace[0.2cm]
& Final			&		&		&		&	2.002604 	\\
& Uncertainty   			&		&		&		&	0.000055 	\\ \hline\hline
	\end{tabular}}
\end{table*}

The value of $g_J$ obtained using different GAS models are summarized in Table \ref{tab:mrci}. First of all, the `e23-SD' and `e33-SD' models are implemented with the dyall.cv3z, dyall.cv4z, diffusion augmented and $spdfg$-densified dyall.cv4z basis sets. All of these calculations have shown the excellent convergence of the obtained $g_J$ values with the increasing basis sets as well as inclusion of more inner core and virtual spinors. However, the results using the `e23-SDT' model shows great changes, which indicates the important role of the triple excitation to the $g_J$ value. The `e23-SDT-I'$<$10 calculation with the dyall.cv4z basis set obtains $g_J=2.002604$. The result of the `e23-SDT-II'$<$10 calculation, which adds SD excitation arising from the $5s^25p^6$ shells, is consistent with the `e23-SDT-I'$<$10 calculation with the same basis set. Repeating the `e23-SDT-I'$<$10 calculation with the d-aug-dyall.cv4z basis set shows the fluctuation of the $g_J$ values around 0.000055. Therefore, we take the $g_J$ value obtained by the `e23-SDT-I'$<$10 calculation with the dyall.cv4z basis set as the final value, while taking the difference between that and one with the d-aug-dyall.cv4z basis set as the uncertainty. To check the dependence of the calculation results on the DHF reference state, we have also conducted the `e23-SD'$<$5 calculation with the dyall.cv3z basis set and the `e23-SDT-I'$<$10 calculation with the dyall.cv4z basis set that are based on the Yb$^+~6s_{1/2}$ open-shell DHF calculation. These calculations are marked by `$^{\ast}$' in Table \ref{tab:mrci}. Their results show that the variation of the $g_J$ due to difference of DHF reference spinors is within the margin of our estimation to the uncertainty. Therefore, the $g_J$ in the MRCI calculation is determined to be $2.002604(55)$, where the number in parentheses indicates the uncertainty.

\section{Results and Discussions}

\begin{table}
\caption{The $g_J$ factors, first-order Zeeman coefficients ($K_Z$, in Hz/$\mu$T), and second-order Zeeman coefficients ($K_0$, in mHz/$\mu{\rm T}^2$) for the ground-state $^{171}$Yb$^+$. Results from previous studies are included for comparison. Abbreviations: MCDHF, multiconfiguration Dirac-Hartree-Fock; MRCI, multireference configuration interaction; RCC, relativistic coupled-cluster; TDHF, time-dependent Hartree-Fock; Nonrel., nonrelativistic; Spectr., spectroscopic measurement.} \label{tab:re}
{\setlength{\tabcolsep}{4pt}
\begin{tabular}{llllllll}\hline\hline
\multicolumn{4}{c}{$g_J$ factor} \\
 \multicolumn{2}{l}{$g_J$} & Source & Type \\\hline
 \multicolumn{2}{l}{2.002626(57)}  & MCDHF & Theor.\\
 \multicolumn{2}{l}{2.002604(55)} & MRCI & Theor.\\
 \multicolumn{2}{l}{\bf 2.002615(70)}  & {\bf Final} & {\bf Theor.}\\
 \multicolumn{2}{l}{2.002798(113)}  & RCC \cite{yu2020ground}  & Theor.\\
 \multicolumn{2}{l}{2.003117}  & TDHF \cite{gossel2013calculation} & Theor. \\
 \multicolumn{2}{l}{2.0023}  & Nonrel. \cite{vanier1989quantum} & Theor. \\
 \multicolumn{2}{l}{1.998}  & Spectr. \cite{meggers1967second} & Exp. \\\\
\multicolumn{4}{c}{Zeeman coefficients} \\
  $K_Z$ & $K_0$  & Source & Type \\\hline
  {\bf 14,010.78(49)} & {\bf 31.0869(22)} &  {\bf Final} & {\bf Theor.}\\
  14,008.6 & 31.077  & Nonrel. \cite{vanier1989quantum} & Theor.\\
  14,012.06(79)$^a$ & 31.0926(35)$^a$  & RCC \cite{yu2020ground}$^a$  & Theor. \\
  14,014.30$^a$ & 31.1025$^a$ & TDHF \cite{gossel2013calculation}$^a$ &Theor.\\
  13,980$^a$ & 30.9$^a$ & Spectr. \cite{meggers1967second}$^a$ & Exp. \\
\hline\hline
\multicolumn{4}{l}{$^a$Derived by using $g_J$ provided in each Refs.} \\
\end{tabular}}
\end{table}

The $g_J$ factor, along with the first- and second-order Zeeman coefficients of the $^{171}$Yb$^+$ ground-state determined in this work, are presented in Table \ref{tab:re}, with uncertainties provided in parentheses. Our MCDHF and MRCI calculations yield \( g_J = 2.002626(57) \) and \( g_J = 2.002604(55) \), respectively. Note that these are two independent, back-to-back calculations, with no computational parameters adjusted to align the two results. Based on the mean value of the MCDHF and MRCI calculated results, we determined \( g_J = 2.002615(70) \), with the value in parentheses representing the upper bound of the uncertainty, covering the uncertainty budgets of both the MCDHF and MRCI calculations \cite{safronova2011precision, bohman2023enhancing, tang2024blackbody}. Another method for evaluating uncertainty based on the weighted mean is presented in the appendix.

The final result, \( g_J = 2.002615(70) \), lies near the lower bound of the previously RCC value, \( g_J = 2.002798(113) \) \cite{yu2020ground}, while improving the consistency of the Land\'{e} \( g_J \) factor to the fifth decimal place, as shown in Table \ref{tab:re}. The TDHF method yields \( g_J = 2.003117 \) \cite{gossel2013calculation}, which is larger than our final result as well as others. The only available experimental result, \( g_J = 1.998 \), comes from an early spectroscopic measurement with only four effective digits and no reported uncertainty \cite{meggers1967second}, and is substantially lower than our result as well as others. This discrepancy highlights the necessity for further measurements. Accurate calculations of the Land\'{e} \( g_J \) factor remain challenging, even for alkali-metal atoms and alkali-metal-like ions \cite{sahoo2017relativistic}, due to their sensitivity to electron correlation effects. These challenges are further compounded in Yb\( ^+ \) by the complexities associated with \( 4f \)-orbital electrons, which have been shown to be both significant and intricate \cite{nandy2014quadrupole}. Thus, the results presented in this work could provide valuable insights into the electron correlation effects contributing to the Land\'{e} \( g_J \) factor of ground-state Yb\( ^+ \).

Further, the values of $K_0$ and $K_Z$ for the $^{171}$Yb$^+$ ground-state are recommended to be $31.0869(22)$ mHz/$\mu$T$^2$ and $14,010.78(49)$ Hz/$\mu$T, respectively, based on the calculated $g_J = 2.002615(70)$, $g'_I = -0.5377\times10^{-3}$ \cite{stone2016table}, and $A = 12,642,812,118$ Hz \cite{han2024determination}, as substituted into Eqs. (\ref{eq:K0}) and (\ref{eq:KZ}). Within the QFS community, the prevailing values for $K_0$ and $K_Z$ for the $^{171}$Yb$^+$ ground-state are $31.077$ mHz/$\mu{\rm T}^2$ and $14,008.6$ Hz/$\mu{\rm T}$, respectively \cite{vanier1989quantum}. However, the coefficients in Ref. \cite{vanier1989quantum} are derived using a non-relativistic $g_J$ of 2.0023 (Nonrel.), which may underestimate the accuracy of the clock transition. The values reported in this work are consistent with the prevailing ones but offer improved reliability. 

Assuming the $B_0$ value for our typical experimental conditions is known precisely, the fractional SOZS $\delta \nu_{\rm SOZS}$ caused by $K_0$ can be calculated to be 
\begin{eqnarray}
\delta \nu_{\rm SOZS}&=& |\frac{1}{A}\frac{\partial \Delta \nu_{\rm SOZS}}{\partial K_0}|~\delta K_0 \
=|\frac{B_0^2}{A}|~\delta K_0\nonumber \\
&<&2\times10^{-18},
\end{eqnarray}
where we have used $B_0 = 0.1$ $\mu$T under ideal experimental conditions \cite{berkeland1998laser}, with $A = 12,642,812,118$ Hz \cite{han2024determination}, $K_0 = 31.0869$ mHz/$\mu{\rm T}^2$, and $\delta K_0 = 0.0022$ mHz/$\mu{\rm T}^2$ as the recommended values for the magnitude and uncertainty of $K_0$, respectively, based on this work. Therefore, the uncertainty in $K_0$ will not impact the SOZS at a level that would compromise the accuracy of QFS operations at the $2\times10^{-18}$ level, provided the applied magnetic field remains below the assumed value.

It is also imperative to calibrate the maximum level of fluctuation of $B_0$. The uncertainty in the calibration of $B_0$ with respect to the $K_Z$ is given by
\begin{eqnarray}
\delta B_0&=&|\frac{\partial B_0}{\partial K_Z}|~\delta K_Z
=|\frac{\Delta \nu_L}{2K_Z^2}|~\delta K_Z\nonumber\\
&<&0.004~{\rm [n T]},
\end{eqnarray}
where we have used $\Delta \nu_L = 2.8$ kHz for $B_0 = 0.1$ nT under ideal experimental conditions \cite{berkeland1998laser}, with $K_Z = 14,010.78$ Hz/$\mu$T and $\delta K_Z = 0.49$ Hz/$\mu$T representing the magnitude and uncertainty of $K_Z$, respectively, based on this work. Accordingly, the calibration of $B_0$ will be affected by the uncertainty in $K_Z$. Using our recommended $K_Z$ value, the uncertainty in $B_0$ is expected to be less than $0.004$ nT, which is sufficiently low to keep the uncertainty in the fractional SOZS relative to the clock frequency below $10^{-17}$ for the assumed magnetic field. The $K_0$ and $K_Z$ factors determined in this work meet the accuracy requirements of state-of-the-art ($1\times10^{-14}$ \cite{phoonthong2014determination}) and our anticipated ($9\times10^{-15}$ \cite{xin2022laser, han2024determination}) microwave QFSs based on $^{171}$Yb$^+$ ions, supporting further advancements in such QFSs.

\section{Conclusion}
In conclusion, we report the determination of the Land\'{e} $g_J$ factor and Zeeman coefficients for the ground-state HFS of $^{171}$Yb$^+$. Systematic calculations of the $g_J$ are performed using two independent methods: MCDHF and MRCI. These calculations yield consistent results, with a $g_J$ factor of 2.002615(70), extending the consistency to the fifth decimal place. Based on this $g_J$ factor, we derive precise values for the first- and second-order Zeeman coefficients, $14,010.78(49)$ Hz/$\mu$T and $31.0869(22)$ mHz/$\mu$T$^2$, respectively. The results presented here support the development of high-accuracy $^{171}$Yb$^+$ microwave QFSs and offer potential for improved constraints on variations in fundamental constants through frequency comparisons involving $^{171}$Yb$^+$ optical and hyperfine transitions, and advancing quantum computers based on trapped-ion.

\section*{Acknowledgments}

This work was supported by the National Key Research and Development Program of China (grant nos. 2021YFA1402100 and 2021YFB2801800), the National Natural Science Foundation of China (grant nos. 12474250 and 11874090) and the Space Application System of China Manned Space Program. We sincerely appreciate the efforts of the editor and the two reviewers in improving the quality of the manuscript. 

J. H. and B. L. contributed equally to this work. 

\section*{APPENDIX}
The formula for calculating the weighted mean \( \bar{x} \) is
\begin{equation}
    \bar{x} = \frac{\sum_{i=1}^{n} w_i x_i}{\sum_{i=1}^{n} w_i},
\end{equation}
where the weight \( w_i = \frac{1}{(\delta x_i)^2} \), representing the weighting of each data point according to its uncertainty. The combined uncertainty of the weighted mean \( \delta \bar{x} \) is calculated as,
\begin{equation}
\delta \bar{x} = \sqrt{\frac{1}{\sum_{i=1}^{n} w_i}}.
\end{equation}
Substituting the MCDHF and MRCI calculation results, 2.002626(57) and 2.002604(55), respectively, the weighted mean result is $2.002615(40)$, where the number in parentheses indicates the weighted mean uncertainty.

\bibliography{apssamp}% Produces the bibliography via BibTeX.

\providecommand{\noopsort}[1]{}\providecommand{\singleletter}[1]{#1}%
\begin{thebibliography}{61}%
\makeatletter
\providecommand \@ifxundefined [1]{%
 \@ifx{#1\undefined}
}%
\providecommand \@ifnum [1]{%
 \ifnum #1\expandafter \@firstoftwo
 \else \expandafter \@secondoftwo
 \fi
}%
\providecommand \@ifx [1]{%
 \ifx #1\expandafter \@firstoftwo
 \else \expandafter \@secondoftwo
 \fi
}%
\providecommand \natexlab [1]{#1}%
\providecommand \enquote  [1]{``#1''}%
\providecommand \bibnamefont  [1]{#1}%
\providecommand \bibfnamefont [1]{#1}%
\providecommand \citenamefont [1]{#1}%
\providecommand \href@noop [0]{\@secondoftwo}%
\providecommand \href [0]{\begingroup \@sanitize@url \@href}%
\providecommand \@href[1]{\@@startlink{#1}\@@href}%
\providecommand \@@href[1]{\endgroup#1\@@endlink}%
\providecommand \@sanitize@url [0]{\catcode `\\12\catcode `\$12\catcode
  `\&12\catcode `\#12\catcode `\^12\catcode `\_12\catcode `\%12\relax}%
\providecommand \@@startlink[1]{}%
\providecommand \@@endlink[0]{}%
\providecommand \url  [0]{\begingroup\@sanitize@url \@url }%
\providecommand \@url [1]{\endgroup\@href {#1}{\urlprefix }}%
\providecommand \urlprefix  [0]{URL }%
\providecommand \Eprint [0]{\href }%
\providecommand \doibase [0]{http://dx.doi.org/}%
\providecommand \selectlanguage [0]{\@gobble}%
\providecommand \bibinfo  [0]{\@secondoftwo}%
\providecommand \bibfield  [0]{\@secondoftwo}%
\providecommand \translation [1]{[#1]}%
\providecommand \BibitemOpen [0]{}%
\providecommand \bibitemStop [0]{}%
\providecommand \bibitemNoStop [0]{.\EOS\space}%
\providecommand \EOS [0]{\spacefactor3000\relax}%
\providecommand \BibitemShut  [1]{\csname bibitem#1\endcsname}%
\let\auto@bib@innerbib\@empty
%</preamble>
\bibitem [{\citenamefont {Schmittberger}\ and\ \citenamefont
  {Scherer}(2020)}]{schmittberger2020review}%
  \BibitemOpen
  \bibfield  {author} {\bibinfo {author} {\bibfnamefont {B.~L.}\ \bibnamefont
  {Schmittberger}}\ and\ \bibinfo {author} {\bibfnamefont {D.~R.}\ \bibnamefont
  {Scherer}},\ }\href@noop {} {\bibfield  {journal} {\bibinfo  {journal} {arXiv
  preprint arXiv:2004.09987}\ } (\bibinfo {year} {2020})}\BibitemShut {NoStop}%
\bibitem [{\citenamefont {Berkeland}\ \emph {et~al.}(1998)\citenamefont
  {Berkeland}, \citenamefont {Miller}, \citenamefont {Bergquist}, \citenamefont
  {Itano},\ and\ \citenamefont {Wineland}}]{berkeland1998laser}%
  \BibitemOpen
  \bibfield  {author} {\bibinfo {author} {\bibfnamefont {D.}~\bibnamefont
  {Berkeland}}, \bibinfo {author} {\bibfnamefont {J.}~\bibnamefont {Miller}},
  \bibinfo {author} {\bibfnamefont {J.~C.}\ \bibnamefont {Bergquist}}, \bibinfo
  {author} {\bibfnamefont {W.~M.}\ \bibnamefont {Itano}}, \ and\ \bibinfo
  {author} {\bibfnamefont {D.~J.}\ \bibnamefont {Wineland}},\ }\href@noop {}
  {\bibfield  {journal} {\bibinfo  {journal} {Physical Review Letters}\
  }\textbf {\bibinfo {volume} {80}},\ \bibinfo {pages} {2089} (\bibinfo {year}
  {1998})}\BibitemShut {NoStop}%
\bibitem [{\citenamefont {Burt}\ \emph {et~al.}(2021)\citenamefont {Burt},
  \citenamefont {Prestage}, \citenamefont {Tjoelker}, \citenamefont {Enzer},
  \citenamefont {Kuang}, \citenamefont {Murphy}, \citenamefont {Robison},
  \citenamefont {Seubert}, \citenamefont {Wang},\ and\ \citenamefont
  {Ely}}]{burt2021demonstration}%
  \BibitemOpen
  \bibfield  {author} {\bibinfo {author} {\bibfnamefont {E.}~\bibnamefont
  {Burt}}, \bibinfo {author} {\bibfnamefont {J.}~\bibnamefont {Prestage}},
  \bibinfo {author} {\bibfnamefont {R.}~\bibnamefont {Tjoelker}}, \bibinfo
  {author} {\bibfnamefont {D.}~\bibnamefont {Enzer}}, \bibinfo {author}
  {\bibfnamefont {D.}~\bibnamefont {Kuang}}, \bibinfo {author} {\bibfnamefont
  {D.}~\bibnamefont {Murphy}}, \bibinfo {author} {\bibfnamefont
  {D.}~\bibnamefont {Robison}}, \bibinfo {author} {\bibfnamefont
  {J.}~\bibnamefont {Seubert}}, \bibinfo {author} {\bibfnamefont
  {R.}~\bibnamefont {Wang}}, \ and\ \bibinfo {author} {\bibfnamefont
  {T.}~\bibnamefont {Ely}},\ }\href@noop {} {\bibfield  {journal} {\bibinfo
  {journal} {Nature}\ }\textbf {\bibinfo {volume} {595}},\ \bibinfo {pages}
  {43} (\bibinfo {year} {2021})}\BibitemShut {NoStop}%
\bibitem [{\citenamefont {Tanaka}\ \emph {et~al.}(1996)\citenamefont {Tanaka},
  \citenamefont {Imajo}, \citenamefont {Hayasaka}, \citenamefont {Ohmukai},
  \citenamefont {Watanabe},\ and\ \citenamefont
  {Urabe}}]{tanaka1996determination}%
  \BibitemOpen
  \bibfield  {author} {\bibinfo {author} {\bibfnamefont {U.}~\bibnamefont
  {Tanaka}}, \bibinfo {author} {\bibfnamefont {H.}~\bibnamefont {Imajo}},
  \bibinfo {author} {\bibfnamefont {K.}~\bibnamefont {Hayasaka}}, \bibinfo
  {author} {\bibfnamefont {R.}~\bibnamefont {Ohmukai}}, \bibinfo {author}
  {\bibfnamefont {M.}~\bibnamefont {Watanabe}}, \ and\ \bibinfo {author}
  {\bibfnamefont {S.}~\bibnamefont {Urabe}},\ }\href@noop {} {\bibfield
  {journal} {\bibinfo  {journal} {Physical Review A}\ }\textbf {\bibinfo
  {volume} {53}},\ \bibinfo {pages} {3982} (\bibinfo {year}
  {1996})}\BibitemShut {NoStop}%
\bibitem [{\citenamefont {Jelenkovi{\'c}}\ \emph {et~al.}(2006)\citenamefont
  {Jelenkovi{\'c}}, \citenamefont {Chung}, \citenamefont {Prestage},\ and\
  \citenamefont {Maleki}}]{jelenkovic2006high}%
  \BibitemOpen
  \bibfield  {author} {\bibinfo {author} {\bibfnamefont {B.}~\bibnamefont
  {Jelenkovi{\'c}}}, \bibinfo {author} {\bibfnamefont {S.}~\bibnamefont
  {Chung}}, \bibinfo {author} {\bibfnamefont {J.}~\bibnamefont {Prestage}}, \
  and\ \bibinfo {author} {\bibfnamefont {L.}~\bibnamefont {Maleki}},\
  }\href@noop {} {\bibfield  {journal} {\bibinfo  {journal} {Physical Review
  A}\ }\textbf {\bibinfo {volume} {74}},\ \bibinfo {pages} {022505} (\bibinfo
  {year} {2006})}\BibitemShut {NoStop}%
\bibitem [{\citenamefont {Han}\ \emph {et~al.}(2019{\natexlab{a}})\citenamefont
  {Han}, \citenamefont {Zuo}, \citenamefont {Zhang},\ and\ \citenamefont
  {Wang}}]{han2019theoretical}%
  \BibitemOpen
  \bibfield  {author} {\bibinfo {author} {\bibfnamefont {J.}~\bibnamefont
  {Han}}, \bibinfo {author} {\bibfnamefont {Y.}~\bibnamefont {Zuo}}, \bibinfo
  {author} {\bibfnamefont {J.}~\bibnamefont {Zhang}}, \ and\ \bibinfo {author}
  {\bibfnamefont {L.}~\bibnamefont {Wang}},\ }\href@noop {} {\bibfield
  {journal} {\bibinfo  {journal} {The European Physical Journal D}\ }\textbf
  {\bibinfo {volume} {73}},\ \bibinfo {pages} {1} (\bibinfo {year}
  {2019}{\natexlab{a}})}\BibitemShut {NoStop}%
\bibitem [{\citenamefont {Han}\ \emph {et~al.}(2019{\natexlab{b}})\citenamefont
  {Han}, \citenamefont {Yu}, \citenamefont {Sahoo}, \citenamefont {Zhang},\
  and\ \citenamefont {Wang}}]{han2019roles}%
  \BibitemOpen
  \bibfield  {author} {\bibinfo {author} {\bibfnamefont {J.}~\bibnamefont
  {Han}}, \bibinfo {author} {\bibfnamefont {Y.}~\bibnamefont {Yu}}, \bibinfo
  {author} {\bibfnamefont {B.}~\bibnamefont {Sahoo}}, \bibinfo {author}
  {\bibfnamefont {J.}~\bibnamefont {Zhang}}, \ and\ \bibinfo {author}
  {\bibfnamefont {L.}~\bibnamefont {Wang}},\ }\href@noop {} {\bibfield
  {journal} {\bibinfo  {journal} {Physical Review A}\ }\textbf {\bibinfo
  {volume} {100}},\ \bibinfo {pages} {042508} (\bibinfo {year}
  {2019}{\natexlab{b}})}\BibitemShut {NoStop}%
\bibitem [{\citenamefont {Han}\ \emph {et~al.}(2021)\citenamefont {Han},
  \citenamefont {Qin}, \citenamefont {Xin}, \citenamefont {Yu}, \citenamefont
  {Dzuba}, \citenamefont {Zhang},\ and\ \citenamefont {Wang}}]{han2021toward}%
  \BibitemOpen
  \bibfield  {author} {\bibinfo {author} {\bibfnamefont {J.}~\bibnamefont
  {Han}}, \bibinfo {author} {\bibfnamefont {H.}~\bibnamefont {Qin}}, \bibinfo
  {author} {\bibfnamefont {N.}~\bibnamefont {Xin}}, \bibinfo {author}
  {\bibfnamefont {Y.}~\bibnamefont {Yu}}, \bibinfo {author} {\bibfnamefont
  {V.}~\bibnamefont {Dzuba}}, \bibinfo {author} {\bibfnamefont
  {J.}~\bibnamefont {Zhang}}, \ and\ \bibinfo {author} {\bibfnamefont
  {L.}~\bibnamefont {Wang}},\ }\href@noop {} {\bibfield  {journal} {\bibinfo
  {journal} {Applied Physics Letters}\ }\textbf {\bibinfo {volume} {118}}
  (\bibinfo {year} {2021})}\BibitemShut {NoStop}%
\bibitem [{\citenamefont {Han}\ \emph {et~al.}(2022{\natexlab{a}})\citenamefont
  {Han}, \citenamefont {Pan}, \citenamefont {Zhang}, \citenamefont {Yang},
  \citenamefont {Zhang}, \citenamefont {Berengut}, \citenamefont {Goriely},
  \citenamefont {Wang}, \citenamefont {Yu}, \citenamefont {Meng} \emph
  {et~al.}}]{han2022isotope}%
  \BibitemOpen
  \bibfield  {author} {\bibinfo {author} {\bibfnamefont {J.}~\bibnamefont
  {Han}}, \bibinfo {author} {\bibfnamefont {C.}~\bibnamefont {Pan}}, \bibinfo
  {author} {\bibfnamefont {K.}~\bibnamefont {Zhang}}, \bibinfo {author}
  {\bibfnamefont {X.}~\bibnamefont {Yang}}, \bibinfo {author} {\bibfnamefont
  {S.}~\bibnamefont {Zhang}}, \bibinfo {author} {\bibfnamefont
  {J.}~\bibnamefont {Berengut}}, \bibinfo {author} {\bibfnamefont
  {S.}~\bibnamefont {Goriely}}, \bibinfo {author} {\bibfnamefont
  {H.}~\bibnamefont {Wang}}, \bibinfo {author} {\bibfnamefont {Y.}~\bibnamefont
  {Yu}}, \bibinfo {author} {\bibfnamefont {J.}~\bibnamefont {Meng}},  \emph
  {et~al.},\ }\href@noop {} {\bibfield  {journal} {\bibinfo  {journal}
  {Physical Review Research}\ }\textbf {\bibinfo {volume} {4}},\ \bibinfo
  {pages} {033049} (\bibinfo {year} {2022}{\natexlab{a}})}\BibitemShut
  {NoStop}%
\bibitem [{\citenamefont {Han}\ \emph {et~al.}(2022{\natexlab{b}})\citenamefont
  {Han}, \citenamefont {Si}, \citenamefont {Qin}, \citenamefont {Xin},
  \citenamefont {Chen}, \citenamefont {Miao}, \citenamefont {Chen},
  \citenamefont {Zhang},\ and\ \citenamefont {Wang}}]{han2022determination}%
  \BibitemOpen
  \bibfield  {author} {\bibinfo {author} {\bibfnamefont {J.}~\bibnamefont
  {Han}}, \bibinfo {author} {\bibfnamefont {R.}~\bibnamefont {Si}}, \bibinfo
  {author} {\bibfnamefont {H.}~\bibnamefont {Qin}}, \bibinfo {author}
  {\bibfnamefont {N.}~\bibnamefont {Xin}}, \bibinfo {author} {\bibfnamefont
  {Y.}~\bibnamefont {Chen}}, \bibinfo {author} {\bibfnamefont {S.}~\bibnamefont
  {Miao}}, \bibinfo {author} {\bibfnamefont {C.}~\bibnamefont {Chen}}, \bibinfo
  {author} {\bibfnamefont {J.}~\bibnamefont {Zhang}}, \ and\ \bibinfo {author}
  {\bibfnamefont {L.}~\bibnamefont {Wang}},\ }\href@noop {} {\bibfield
  {journal} {\bibinfo  {journal} {Physical Review A}\ }\textbf {\bibinfo
  {volume} {106}},\ \bibinfo {pages} {012821} (\bibinfo {year}
  {2022}{\natexlab{b}})}\BibitemShut {NoStop}%
\bibitem [{\citenamefont {Qin}\ \emph {et~al.}(2022)\citenamefont {Qin},
  \citenamefont {Miao}, \citenamefont {Han}, \citenamefont {Xin}, \citenamefont
  {Chen}, \citenamefont {Zhang}, \citenamefont {Wang} \emph
  {et~al.}}]{qin2022high}%
  \BibitemOpen
  \bibfield  {author} {\bibinfo {author} {\bibfnamefont {H.-R.}\ \bibnamefont
  {Qin}}, \bibinfo {author} {\bibfnamefont {S.-N.}\ \bibnamefont {Miao}},
  \bibinfo {author} {\bibfnamefont {J.-Z.}\ \bibnamefont {Han}}, \bibinfo
  {author} {\bibfnamefont {N.-C.}\ \bibnamefont {Xin}}, \bibinfo {author}
  {\bibfnamefont {Y.-T.}\ \bibnamefont {Chen}}, \bibinfo {author}
  {\bibfnamefont {J.}~\bibnamefont {Zhang}}, \bibinfo {author} {\bibfnamefont
  {L.}~\bibnamefont {Wang}},  \emph {et~al.},\ }\href@noop {} {\bibfield
  {journal} {\bibinfo  {journal} {Physical Review Applied}\ }\textbf {\bibinfo
  {volume} {18}},\ \bibinfo {pages} {024023} (\bibinfo {year}
  {2022})}\BibitemShut {NoStop}%
\bibitem [{\citenamefont {Park}\ \emph {et~al.}(2007)\citenamefont {Park},
  \citenamefont {Manson}, \citenamefont {Wouters}, \citenamefont {Warrington},
  \citenamefont {Lawn},\ and\ \citenamefont {Fisk}}]{park2007171Yb+}%
  \BibitemOpen
  \bibfield  {author} {\bibinfo {author} {\bibfnamefont {S.~J.}\ \bibnamefont
  {Park}}, \bibinfo {author} {\bibfnamefont {P.~J.}\ \bibnamefont {Manson}},
  \bibinfo {author} {\bibfnamefont {M.~J.}\ \bibnamefont {Wouters}}, \bibinfo
  {author} {\bibfnamefont {R.~B.}\ \bibnamefont {Warrington}}, \bibinfo
  {author} {\bibfnamefont {M.~A.}\ \bibnamefont {Lawn}}, \ and\ \bibinfo
  {author} {\bibfnamefont {P.~T.~H.}\ \bibnamefont {Fisk}},\ }\href {\doibase
  10.1109/FREQ.2007.4319145} {\bibfield  {journal} {\bibinfo  {journal} {2007
  IEEE International Frequency Control Symposium Joint with the 21st European
  Frequency and Time Forum}\ ,\ \bibinfo {pages} {613}} (\bibinfo {year}
  {2007})}\BibitemShut {NoStop}%
\bibitem [{\citenamefont {Phoonthong}\ \emph {et~al.}(2014)\citenamefont
  {Phoonthong}, \citenamefont {Mizuno}, \citenamefont {Kido},\ and\
  \citenamefont {Shiga}}]{phoonthong2014determination}%
  \BibitemOpen
  \bibfield  {author} {\bibinfo {author} {\bibfnamefont {P.}~\bibnamefont
  {Phoonthong}}, \bibinfo {author} {\bibfnamefont {M.}~\bibnamefont {Mizuno}},
  \bibinfo {author} {\bibfnamefont {K.}~\bibnamefont {Kido}}, \ and\ \bibinfo
  {author} {\bibfnamefont {N.}~\bibnamefont {Shiga}},\ }\href@noop {}
  {\bibfield  {journal} {\bibinfo  {journal} {Applied Physics B}\ }\textbf
  {\bibinfo {volume} {117}},\ \bibinfo {pages} {673} (\bibinfo {year}
  {2014})}\BibitemShut {NoStop}%
\bibitem [{\citenamefont {Schwindt}\ \emph {et~al.}(2015)\citenamefont
  {Schwindt}, \citenamefont {Jau}, \citenamefont {Partner}, \citenamefont
  {Serkland}, \citenamefont {Ison}, \citenamefont {McCants}, \citenamefont
  {Winrow}, \citenamefont {Prestage}, \citenamefont {Kellogg}, \citenamefont
  {Yu} \emph {et~al.}}]{schwindt2015miniature}%
  \BibitemOpen
  \bibfield  {author} {\bibinfo {author} {\bibfnamefont {P.~D.}\ \bibnamefont
  {Schwindt}}, \bibinfo {author} {\bibfnamefont {Y.-Y.}\ \bibnamefont {Jau}},
  \bibinfo {author} {\bibfnamefont {H.~L.}\ \bibnamefont {Partner}}, \bibinfo
  {author} {\bibfnamefont {D.~K.}\ \bibnamefont {Serkland}}, \bibinfo {author}
  {\bibfnamefont {A.}~\bibnamefont {Ison}}, \bibinfo {author} {\bibfnamefont
  {A.}~\bibnamefont {McCants}}, \bibinfo {author} {\bibfnamefont
  {E.}~\bibnamefont {Winrow}}, \bibinfo {author} {\bibfnamefont
  {J.}~\bibnamefont {Prestage}}, \bibinfo {author} {\bibfnamefont
  {J.}~\bibnamefont {Kellogg}}, \bibinfo {author} {\bibfnamefont
  {N.}~\bibnamefont {Yu}},  \emph {et~al.},\ }\href@noop {} {\bibfield
  {journal} {\bibinfo  {journal} {2015 Joint Conference of the IEEE
  International Frequency Control Symposium \& the European Frequency and Time
  Forum}\ ,\ \bibinfo {pages} {752}} (\bibinfo {year} {2015})}\BibitemShut
  {NoStop}%
\bibitem [{\citenamefont {Schwindt}\ \emph {et~al.}(2016)\citenamefont
  {Schwindt}, \citenamefont {Jau}, \citenamefont {Partner}, \citenamefont
  {Casias}, \citenamefont {Wagner}, \citenamefont {Moorman}, \citenamefont
  {Manginell}, \citenamefont {Kellogg},\ and\ \citenamefont
  {Prestage}}]{schwindt2016highly}%
  \BibitemOpen
  \bibfield  {author} {\bibinfo {author} {\bibfnamefont {P.~D.}\ \bibnamefont
  {Schwindt}}, \bibinfo {author} {\bibfnamefont {Y.-Y.}\ \bibnamefont {Jau}},
  \bibinfo {author} {\bibfnamefont {H.}~\bibnamefont {Partner}}, \bibinfo
  {author} {\bibfnamefont {A.}~\bibnamefont {Casias}}, \bibinfo {author}
  {\bibfnamefont {A.~R.}\ \bibnamefont {Wagner}}, \bibinfo {author}
  {\bibfnamefont {M.}~\bibnamefont {Moorman}}, \bibinfo {author} {\bibfnamefont
  {R.~P.}\ \bibnamefont {Manginell}}, \bibinfo {author} {\bibfnamefont {J.~R.}\
  \bibnamefont {Kellogg}}, \ and\ \bibinfo {author} {\bibfnamefont {J.~D.}\
  \bibnamefont {Prestage}},\ }\href@noop {} {\bibfield  {journal} {\bibinfo
  {journal} {Review of Scientific Instruments}\ }\textbf {\bibinfo {volume}
  {87}} (\bibinfo {year} {2016})}\BibitemShut {NoStop}%
\bibitem [{\citenamefont {Mulholland}\ \emph
  {et~al.}(2019{\natexlab{a}})\citenamefont {Mulholland}, \citenamefont
  {Klein}, \citenamefont {Barwood}, \citenamefont {Donnellan}, \citenamefont
  {Nisbet-Jones}, \citenamefont {Huang}, \citenamefont {Walsh}, \citenamefont
  {Baird},\ and\ \citenamefont {Gill}}]{mulholland2019compact}%
  \BibitemOpen
  \bibfield  {author} {\bibinfo {author} {\bibfnamefont {S.}~\bibnamefont
  {Mulholland}}, \bibinfo {author} {\bibfnamefont {H.}~\bibnamefont {Klein}},
  \bibinfo {author} {\bibfnamefont {G.}~\bibnamefont {Barwood}}, \bibinfo
  {author} {\bibfnamefont {S.}~\bibnamefont {Donnellan}}, \bibinfo {author}
  {\bibfnamefont {P.}~\bibnamefont {Nisbet-Jones}}, \bibinfo {author}
  {\bibfnamefont {G.}~\bibnamefont {Huang}}, \bibinfo {author} {\bibfnamefont
  {G.}~\bibnamefont {Walsh}}, \bibinfo {author} {\bibfnamefont
  {P.}~\bibnamefont {Baird}}, \ and\ \bibinfo {author} {\bibfnamefont
  {P.}~\bibnamefont {Gill}},\ }\href@noop {} {\bibfield  {journal} {\bibinfo
  {journal} {Review of Scientific Instruments}\ }\textbf {\bibinfo {volume}
  {90}} (\bibinfo {year} {2019}{\natexlab{a}})}\BibitemShut {NoStop}%
\bibitem [{\citenamefont {Mulholland}\ \emph
  {et~al.}(2019{\natexlab{b}})\citenamefont {Mulholland}, \citenamefont
  {Klein}, \citenamefont {Barwood}, \citenamefont {Donnellan}, \citenamefont
  {Gentle}, \citenamefont {Huang}, \citenamefont {Walsh}, \citenamefont
  {Baird},\ and\ \citenamefont {Gill}}]{mulholland2019laser}%
  \BibitemOpen
  \bibfield  {author} {\bibinfo {author} {\bibfnamefont {S.}~\bibnamefont
  {Mulholland}}, \bibinfo {author} {\bibfnamefont {H.}~\bibnamefont {Klein}},
  \bibinfo {author} {\bibfnamefont {G.}~\bibnamefont {Barwood}}, \bibinfo
  {author} {\bibfnamefont {S.}~\bibnamefont {Donnellan}}, \bibinfo {author}
  {\bibfnamefont {D.}~\bibnamefont {Gentle}}, \bibinfo {author} {\bibfnamefont
  {G.}~\bibnamefont {Huang}}, \bibinfo {author} {\bibfnamefont
  {G.}~\bibnamefont {Walsh}}, \bibinfo {author} {\bibfnamefont
  {P.}~\bibnamefont {Baird}}, \ and\ \bibinfo {author} {\bibfnamefont
  {P.}~\bibnamefont {Gill}},\ }\href@noop {} {\bibfield  {journal} {\bibinfo
  {journal} {Applied Physics B}\ }\textbf {\bibinfo {volume} {125}},\ \bibinfo
  {pages} {198} (\bibinfo {year} {2019}{\natexlab{b}})}\BibitemShut {NoStop}%
\bibitem [{\citenamefont {Han}\ \emph {et~al.}(2023)\citenamefont {Han},
  \citenamefont {Zheng}, \citenamefont {Miao}, \citenamefont {Chen},
  \citenamefont {Zhang}, \citenamefont {Wang}, \citenamefont {Han},
  \citenamefont {Chen}, \citenamefont {Xue},\ and\ \citenamefont
  {Zhang}}]{han2023cooling}%
  \BibitemOpen
  \bibfield  {author} {\bibinfo {author} {\bibfnamefont {J.~Z.}\ \bibnamefont
  {Han}}, \bibinfo {author} {\bibfnamefont {Y.}~\bibnamefont {Zheng}}, \bibinfo
  {author} {\bibfnamefont {S.~N.}\ \bibnamefont {Miao}}, \bibinfo {author}
  {\bibfnamefont {Y.~T.}\ \bibnamefont {Chen}}, \bibinfo {author}
  {\bibfnamefont {J.~W.}\ \bibnamefont {Zhang}}, \bibinfo {author}
  {\bibfnamefont {L.~J.}\ \bibnamefont {Wang}}, \bibinfo {author}
  {\bibfnamefont {L.}~\bibnamefont {Han}}, \bibinfo {author} {\bibfnamefont
  {X.}~\bibnamefont {Chen}}, \bibinfo {author} {\bibfnamefont {X.~B.}\
  \bibnamefont {Xue}}, \ and\ \bibinfo {author} {\bibfnamefont {S.~K.}\
  \bibnamefont {Zhang}},\ }\href@noop {} {\bibfield  {journal} {\bibinfo
  {journal} {Joint Conference of the European Frequency and Time Forum and IEEE
  International Frequency Control Symposium}\ } (\bibinfo {year}
  {2023})}\BibitemShut {NoStop}%
\bibitem [{\citenamefont {Han}\ \emph {et~al.}(2024)\citenamefont {Han},
  \citenamefont {Xin}, \citenamefont {Zhang}, \citenamefont {Yu}, \citenamefont
  {Li}, \citenamefont {Qian},\ and\ \citenamefont
  {Wang}}]{han2024determination}%
  \BibitemOpen
  \bibfield  {author} {\bibinfo {author} {\bibfnamefont {J.}~\bibnamefont
  {Han}}, \bibinfo {author} {\bibfnamefont {N.}~\bibnamefont {Xin}}, \bibinfo
  {author} {\bibfnamefont {J.}~\bibnamefont {Zhang}}, \bibinfo {author}
  {\bibfnamefont {Y.}~\bibnamefont {Yu}}, \bibinfo {author} {\bibfnamefont
  {J.}~\bibnamefont {Li}}, \bibinfo {author} {\bibfnamefont {L.}~\bibnamefont
  {Qian}}, \ and\ \bibinfo {author} {\bibfnamefont {L.}~\bibnamefont {Wang}},\
  }\href@noop {} {\bibfield  {journal} {\bibinfo  {journal} {Applied Physics
  Letters}\ }\textbf {\bibinfo {volume} {125}} (\bibinfo {year}
  {2024})}\BibitemShut {NoStop}%
\bibitem [{\citenamefont {Miao}\ \emph {et~al.}(2021)\citenamefont {Miao},
  \citenamefont {Zhang}, \citenamefont {Qin}, \citenamefont {Xin},
  \citenamefont {Han},\ and\ \citenamefont {Wang}}]{miao2021precision}%
  \BibitemOpen
  \bibfield  {author} {\bibinfo {author} {\bibfnamefont {S.}~\bibnamefont
  {Miao}}, \bibinfo {author} {\bibfnamefont {J.}~\bibnamefont {Zhang}},
  \bibinfo {author} {\bibfnamefont {H.}~\bibnamefont {Qin}}, \bibinfo {author}
  {\bibfnamefont {N.}~\bibnamefont {Xin}}, \bibinfo {author} {\bibfnamefont
  {J.}~\bibnamefont {Han}}, \ and\ \bibinfo {author} {\bibfnamefont
  {L.}~\bibnamefont {Wang}},\ }\href@noop {} {\bibfield  {journal} {\bibinfo
  {journal} {Optics Letters}\ }\textbf {\bibinfo {volume} {46}},\ \bibinfo
  {pages} {5882} (\bibinfo {year} {2021})}\BibitemShut {NoStop}%
\bibitem [{\citenamefont {Warrington}\ \emph {et~al.}(2002)\citenamefont
  {Warrington}, \citenamefont {Fisk}, \citenamefont {Wouters},\ and\
  \citenamefont {Lawn}}]{warrington2002microwave}%
  \BibitemOpen
  \bibfield  {author} {\bibinfo {author} {\bibfnamefont {R.}~\bibnamefont
  {Warrington}}, \bibinfo {author} {\bibfnamefont {P.}~\bibnamefont {Fisk}},
  \bibinfo {author} {\bibfnamefont {M.}~\bibnamefont {Wouters}}, \ and\
  \bibinfo {author} {\bibfnamefont {M.}~\bibnamefont {Lawn}},\ }\href@noop {}
  {\bibfield  {journal} {\bibinfo  {journal} {Frequency Standards and
  Metrology}\ ,\ \bibinfo {pages} {297}} (\bibinfo {year} {2002})}\BibitemShut
  {NoStop}%
\bibitem [{\citenamefont {Hosaka}\ \emph {et~al.}(2005)\citenamefont {Hosaka},
  \citenamefont {Webster}, \citenamefont {Blythe}, \citenamefont {Stannard},
  \citenamefont {Beaton}, \citenamefont {Margolis}, \citenamefont {Lea},\ and\
  \citenamefont {Gill}}]{hosaka2005optical}%
  \BibitemOpen
  \bibfield  {author} {\bibinfo {author} {\bibfnamefont {K.}~\bibnamefont
  {Hosaka}}, \bibinfo {author} {\bibfnamefont {S.~A.}\ \bibnamefont {Webster}},
  \bibinfo {author} {\bibfnamefont {P.~J.}\ \bibnamefont {Blythe}}, \bibinfo
  {author} {\bibfnamefont {A.}~\bibnamefont {Stannard}}, \bibinfo {author}
  {\bibfnamefont {D.}~\bibnamefont {Beaton}}, \bibinfo {author} {\bibfnamefont
  {H.~S.}\ \bibnamefont {Margolis}}, \bibinfo {author} {\bibfnamefont {S.~N.}\
  \bibnamefont {Lea}}, \ and\ \bibinfo {author} {\bibfnamefont
  {P.}~\bibnamefont {Gill}},\ }\href@noop {} {\bibfield  {journal} {\bibinfo
  {journal} {IEEE Transactions on Instrumentation and Measurement}\ }\textbf
  {\bibinfo {volume} {54}},\ \bibinfo {pages} {759} (\bibinfo {year}
  {2005})}\BibitemShut {NoStop}%
\bibitem [{\citenamefont {Rosenband}\ \emph {et~al.}(2008)\citenamefont
  {Rosenband}, \citenamefont {Hume}, \citenamefont {Schmidt}, \citenamefont
  {Chou}, \citenamefont {Brusch}, \citenamefont {Lorini}, \citenamefont
  {Oskay}, \citenamefont {Drullinger}, \citenamefont {Fortier}, \citenamefont
  {Stalnaker} \emph {et~al.}}]{rosenband2008frequency}%
  \BibitemOpen
  \bibfield  {author} {\bibinfo {author} {\bibfnamefont {T.}~\bibnamefont
  {Rosenband}}, \bibinfo {author} {\bibfnamefont {D.}~\bibnamefont {Hume}},
  \bibinfo {author} {\bibfnamefont {P.}~\bibnamefont {Schmidt}}, \bibinfo
  {author} {\bibfnamefont {C.-W.}\ \bibnamefont {Chou}}, \bibinfo {author}
  {\bibfnamefont {A.}~\bibnamefont {Brusch}}, \bibinfo {author} {\bibfnamefont
  {L.}~\bibnamefont {Lorini}}, \bibinfo {author} {\bibfnamefont
  {W.}~\bibnamefont {Oskay}}, \bibinfo {author} {\bibfnamefont {R.~E.}\
  \bibnamefont {Drullinger}}, \bibinfo {author} {\bibfnamefont {T.~M.}\
  \bibnamefont {Fortier}}, \bibinfo {author} {\bibfnamefont {J.~E.}\
  \bibnamefont {Stalnaker}},  \emph {et~al.},\ }\href@noop {} {\bibfield
  {journal} {\bibinfo  {journal} {Science}\ }\textbf {\bibinfo {volume}
  {319}},\ \bibinfo {pages} {1808} (\bibinfo {year} {2008})}\BibitemShut
  {NoStop}%
\bibitem [{\citenamefont {Godun}\ \emph {et~al.}(2014)\citenamefont {Godun},
  \citenamefont {Nisbet-Jones}, \citenamefont {Jones}, \citenamefont {King},
  \citenamefont {Johnson}, \citenamefont {Margolis}, \citenamefont {Szymaniec},
  \citenamefont {Lea}, \citenamefont {Bongs},\ and\ \citenamefont
  {Gill}}]{godun2014frequency}%
  \BibitemOpen
  \bibfield  {author} {\bibinfo {author} {\bibfnamefont {R.}~\bibnamefont
  {Godun}}, \bibinfo {author} {\bibfnamefont {P.}~\bibnamefont {Nisbet-Jones}},
  \bibinfo {author} {\bibfnamefont {J.}~\bibnamefont {Jones}}, \bibinfo
  {author} {\bibfnamefont {S.}~\bibnamefont {King}}, \bibinfo {author}
  {\bibfnamefont {L.}~\bibnamefont {Johnson}}, \bibinfo {author} {\bibfnamefont
  {H.}~\bibnamefont {Margolis}}, \bibinfo {author} {\bibfnamefont
  {K.}~\bibnamefont {Szymaniec}}, \bibinfo {author} {\bibfnamefont
  {S.}~\bibnamefont {Lea}}, \bibinfo {author} {\bibfnamefont {K.}~\bibnamefont
  {Bongs}}, \ and\ \bibinfo {author} {\bibfnamefont {P.}~\bibnamefont {Gill}},\
  }\href@noop {} {\bibfield  {journal} {\bibinfo  {journal} {Physical Review
  Letters}\ }\textbf {\bibinfo {volume} {113}},\ \bibinfo {pages} {210801}
  (\bibinfo {year} {2014})}\BibitemShut {NoStop}%
\bibitem [{\citenamefont {Brewer}\ \emph {et~al.}(2019)\citenamefont {Brewer},
  \citenamefont {Chen}, \citenamefont {Hankin}, \citenamefont {Clements},
  \citenamefont {Chou}, \citenamefont {Wineland}, \citenamefont {Hume},\ and\
  \citenamefont {Leibrandt}}]{brewer2019al+}%
  \BibitemOpen
  \bibfield  {author} {\bibinfo {author} {\bibfnamefont {S.~M.}\ \bibnamefont
  {Brewer}}, \bibinfo {author} {\bibfnamefont {J.-S.}\ \bibnamefont {Chen}},
  \bibinfo {author} {\bibfnamefont {A.~M.}\ \bibnamefont {Hankin}}, \bibinfo
  {author} {\bibfnamefont {E.~R.}\ \bibnamefont {Clements}}, \bibinfo {author}
  {\bibfnamefont {C.-w.}\ \bibnamefont {Chou}}, \bibinfo {author}
  {\bibfnamefont {D.~J.}\ \bibnamefont {Wineland}}, \bibinfo {author}
  {\bibfnamefont {D.~B.}\ \bibnamefont {Hume}}, \ and\ \bibinfo {author}
  {\bibfnamefont {D.~R.}\ \bibnamefont {Leibrandt}},\ }\href@noop {} {\bibfield
   {journal} {\bibinfo  {journal} {Physical Review Letters}\ }\textbf {\bibinfo
  {volume} {123}},\ \bibinfo {pages} {033201} (\bibinfo {year}
  {2019})}\BibitemShut {NoStop}%
\bibitem [{\citenamefont {Stone}(2019)}]{stone2019table}%
  \BibitemOpen
  \bibfield  {author} {\bibinfo {author} {\bibfnamefont {N.}~\bibnamefont
  {Stone}},\ }\href@noop {} {\bibfield  {journal} {\bibinfo  {journal}
  {International Atomic Energy Agency}\ } (\bibinfo {year} {2019})}\BibitemShut
  {NoStop}%
\bibitem [{\citenamefont {Meggers}(1967)}]{meggers1967second}%
  \BibitemOpen
  \bibfield  {author} {\bibinfo {author} {\bibfnamefont {W.~F.}\ \bibnamefont
  {Meggers}},\ }\href@noop {} {\bibfield  {journal} {\bibinfo  {journal}
  {Journal of Research of the National Bureau of Standards. Section A, Physics
  and Chemistry}\ }\textbf {\bibinfo {volume} {71}},\ \bibinfo {pages} {396}
  (\bibinfo {year} {1967})}\BibitemShut {NoStop}%
\bibitem [{\citenamefont {Yu}\ \emph {et~al.}(2020)\citenamefont {Yu},
  \citenamefont {Sahoo},\ and\ \citenamefont {Suo}}]{yu2020ground}%
  \BibitemOpen
  \bibfield  {author} {\bibinfo {author} {\bibfnamefont {Y.}~\bibnamefont
  {Yu}}, \bibinfo {author} {\bibfnamefont {B.}~\bibnamefont {Sahoo}}, \ and\
  \bibinfo {author} {\bibfnamefont {B.}~\bibnamefont {Suo}},\ }\href@noop {}
  {\bibfield  {journal} {\bibinfo  {journal} {Physical Review A}\ }\textbf
  {\bibinfo {volume} {102}},\ \bibinfo {pages} {062824} (\bibinfo {year}
  {2020})}\BibitemShut {NoStop}%
\bibitem [{\citenamefont {Gossel}\ \emph {et~al.}(2013)\citenamefont {Gossel},
  \citenamefont {Dzuba},\ and\ \citenamefont
  {Flambaum}}]{gossel2013calculation}%
  \BibitemOpen
  \bibfield  {author} {\bibinfo {author} {\bibfnamefont {G.}~\bibnamefont
  {Gossel}}, \bibinfo {author} {\bibfnamefont {V.}~\bibnamefont {Dzuba}}, \
  and\ \bibinfo {author} {\bibfnamefont {V.}~\bibnamefont {Flambaum}},\
  }\href@noop {} {\bibfield  {journal} {\bibinfo  {journal} {Physical Review
  A}\ }\textbf {\bibinfo {volume} {88}},\ \bibinfo {pages} {034501} (\bibinfo
  {year} {2013})}\BibitemShut {NoStop}%
\bibitem [{\citenamefont {Xin}\ \emph {et~al.}(2022)\citenamefont {Xin},
  \citenamefont {Qin}, \citenamefont {Miao}, \citenamefont {Chen},
  \citenamefont {Zheng}, \citenamefont {Han}, \citenamefont {Zhang},\ and\
  \citenamefont {Wang}}]{xin2022laser}%
  \BibitemOpen
  \bibfield  {author} {\bibinfo {author} {\bibfnamefont {N.}~\bibnamefont
  {Xin}}, \bibinfo {author} {\bibfnamefont {H.}~\bibnamefont {Qin}}, \bibinfo
  {author} {\bibfnamefont {S.}~\bibnamefont {Miao}}, \bibinfo {author}
  {\bibfnamefont {Y.}~\bibnamefont {Chen}}, \bibinfo {author} {\bibfnamefont
  {Y.}~\bibnamefont {Zheng}}, \bibinfo {author} {\bibfnamefont
  {J.}~\bibnamefont {Han}}, \bibinfo {author} {\bibfnamefont {J.}~\bibnamefont
  {Zhang}}, \ and\ \bibinfo {author} {\bibfnamefont {L.}~\bibnamefont {Wang}},\
  }\href@noop {} {\bibfield  {journal} {\bibinfo  {journal} {Optics Express}\
  }\textbf {\bibinfo {volume} {30}},\ \bibinfo {pages} {14574} (\bibinfo {year}
  {2022})}\BibitemShut {NoStop}%
\bibitem [{\citenamefont {Lange}\ \emph {et~al.}(2021)\citenamefont {Lange},
  \citenamefont {Huntemann}, \citenamefont {Rahm}, \citenamefont {Sanner},
  \citenamefont {Shao}, \citenamefont {Lipphardt}, \citenamefont {Tamm},
  \citenamefont {Weyers},\ and\ \citenamefont {Peik}}]{lange2021improved}%
  \BibitemOpen
  \bibfield  {author} {\bibinfo {author} {\bibfnamefont {R.}~\bibnamefont
  {Lange}}, \bibinfo {author} {\bibfnamefont {N.}~\bibnamefont {Huntemann}},
  \bibinfo {author} {\bibfnamefont {J.}~\bibnamefont {Rahm}}, \bibinfo {author}
  {\bibfnamefont {C.}~\bibnamefont {Sanner}}, \bibinfo {author} {\bibfnamefont
  {H.}~\bibnamefont {Shao}}, \bibinfo {author} {\bibfnamefont {B.}~\bibnamefont
  {Lipphardt}}, \bibinfo {author} {\bibfnamefont {C.}~\bibnamefont {Tamm}},
  \bibinfo {author} {\bibfnamefont {S.}~\bibnamefont {Weyers}}, \ and\ \bibinfo
  {author} {\bibfnamefont {E.}~\bibnamefont {Peik}},\ }\href@noop {} {\bibfield
   {journal} {\bibinfo  {journal} {Physical Review Letters}\ }\textbf {\bibinfo
  {volume} {126}},\ \bibinfo {pages} {011102} (\bibinfo {year}
  {2021})}\BibitemShut {NoStop}%
\bibitem [{\citenamefont {Filzinger}\ \emph {et~al.}(2023)\citenamefont
  {Filzinger}, \citenamefont {D{\"o}rscher}, \citenamefont {Lange},
  \citenamefont {Klose}, \citenamefont {Steinel}, \citenamefont {Benkler},
  \citenamefont {Peik}, \citenamefont {Lisdat},\ and\ \citenamefont
  {Huntemann}}]{filzinger2023improved}%
  \BibitemOpen
  \bibfield  {author} {\bibinfo {author} {\bibfnamefont {M.}~\bibnamefont
  {Filzinger}}, \bibinfo {author} {\bibfnamefont {S.}~\bibnamefont
  {D{\"o}rscher}}, \bibinfo {author} {\bibfnamefont {R.}~\bibnamefont {Lange}},
  \bibinfo {author} {\bibfnamefont {J.}~\bibnamefont {Klose}}, \bibinfo
  {author} {\bibfnamefont {M.}~\bibnamefont {Steinel}}, \bibinfo {author}
  {\bibfnamefont {E.}~\bibnamefont {Benkler}}, \bibinfo {author} {\bibfnamefont
  {E.}~\bibnamefont {Peik}}, \bibinfo {author} {\bibfnamefont {C.}~\bibnamefont
  {Lisdat}}, \ and\ \bibinfo {author} {\bibfnamefont {N.}~\bibnamefont
  {Huntemann}},\ }\href@noop {} {\bibfield  {journal} {\bibinfo  {journal}
  {Physical Review Letters}\ }\textbf {\bibinfo {volume} {130}},\ \bibinfo
  {pages} {253001} (\bibinfo {year} {2023})}\BibitemShut {NoStop}%
\bibitem [{\citenamefont {Flambaum}\ and\ \citenamefont
  {Tedesco}(2006)}]{flambaum2006dependence}%
  \BibitemOpen
  \bibfield  {author} {\bibinfo {author} {\bibfnamefont {V.}~\bibnamefont
  {Flambaum}}\ and\ \bibinfo {author} {\bibfnamefont {A.}~\bibnamefont
  {Tedesco}},\ }\href@noop {} {\bibfield  {journal} {\bibinfo  {journal}
  {Physical Review C}\ }\textbf {\bibinfo {volume} {73}},\ \bibinfo {pages}
  {055501} (\bibinfo {year} {2006})}\BibitemShut {NoStop}%
\bibitem [{\citenamefont {Dinh}\ \emph {et~al.}(2009)\citenamefont {Dinh},
  \citenamefont {Dunning}, \citenamefont {Dzuba},\ and\ \citenamefont
  {Flambaum}}]{dinh2009sensitivity}%
  \BibitemOpen
  \bibfield  {author} {\bibinfo {author} {\bibfnamefont {T.}~\bibnamefont
  {Dinh}}, \bibinfo {author} {\bibfnamefont {A.}~\bibnamefont {Dunning}},
  \bibinfo {author} {\bibfnamefont {V.}~\bibnamefont {Dzuba}}, \ and\ \bibinfo
  {author} {\bibfnamefont {V.}~\bibnamefont {Flambaum}},\ }\href@noop {}
  {\bibfield  {journal} {\bibinfo  {journal} {Physical Review A}\ }\textbf
  {\bibinfo {volume} {79}},\ \bibinfo {pages} {054102} (\bibinfo {year}
  {2009})}\BibitemShut {NoStop}%
\bibitem [{\citenamefont {Pino}\ \emph {et~al.}(2021)\citenamefont {Pino},
  \citenamefont {Dreiling}, \citenamefont {Figgatt}, \citenamefont {Gaebler},
  \citenamefont {Moses}, \citenamefont {Allman}, \citenamefont {Baldwin},
  \citenamefont {Foss-Feig}, \citenamefont {Hayes}, \citenamefont {Mayer} \emph
  {et~al.}}]{pino2021demonstration}%
  \BibitemOpen
  \bibfield  {author} {\bibinfo {author} {\bibfnamefont {J.~M.}\ \bibnamefont
  {Pino}}, \bibinfo {author} {\bibfnamefont {J.~M.}\ \bibnamefont {Dreiling}},
  \bibinfo {author} {\bibfnamefont {C.}~\bibnamefont {Figgatt}}, \bibinfo
  {author} {\bibfnamefont {J.~P.}\ \bibnamefont {Gaebler}}, \bibinfo {author}
  {\bibfnamefont {S.~A.}\ \bibnamefont {Moses}}, \bibinfo {author}
  {\bibfnamefont {M.}~\bibnamefont {Allman}}, \bibinfo {author} {\bibfnamefont
  {C.}~\bibnamefont {Baldwin}}, \bibinfo {author} {\bibfnamefont
  {M.}~\bibnamefont {Foss-Feig}}, \bibinfo {author} {\bibfnamefont
  {D.}~\bibnamefont {Hayes}}, \bibinfo {author} {\bibfnamefont
  {K.}~\bibnamefont {Mayer}},  \emph {et~al.},\ }\href@noop {} {\bibfield
  {journal} {\bibinfo  {journal} {Nature}\ }\textbf {\bibinfo {volume} {592}},\
  \bibinfo {pages} {209} (\bibinfo {year} {2021})}\BibitemShut {NoStop}%
\bibitem [{\citenamefont {Feng}\ \emph {et~al.}(2023)\citenamefont {Feng},
  \citenamefont {Katz}, \citenamefont {Haack}, \citenamefont {Maghrebi},
  \citenamefont {Gorshkov}, \citenamefont {Gong}, \citenamefont {Cetina},\ and\
  \citenamefont {Monroe}}]{feng2023continuous}%
  \BibitemOpen
  \bibfield  {author} {\bibinfo {author} {\bibfnamefont {L.}~\bibnamefont
  {Feng}}, \bibinfo {author} {\bibfnamefont {O.}~\bibnamefont {Katz}}, \bibinfo
  {author} {\bibfnamefont {C.}~\bibnamefont {Haack}}, \bibinfo {author}
  {\bibfnamefont {M.}~\bibnamefont {Maghrebi}}, \bibinfo {author}
  {\bibfnamefont {A.~V.}\ \bibnamefont {Gorshkov}}, \bibinfo {author}
  {\bibfnamefont {Z.}~\bibnamefont {Gong}}, \bibinfo {author} {\bibfnamefont
  {M.}~\bibnamefont {Cetina}}, \ and\ \bibinfo {author} {\bibfnamefont
  {C.}~\bibnamefont {Monroe}},\ }\href@noop {} {\bibfield  {journal} {\bibinfo
  {journal} {Nature}\ }\textbf {\bibinfo {volume} {623}},\ \bibinfo {pages}
  {713} (\bibinfo {year} {2023})}\BibitemShut {NoStop}%
\bibitem [{\citenamefont {Guo}\ \emph {et~al.}(2024)\citenamefont {Guo},
  \citenamefont {Wu}, \citenamefont {Ye}, \citenamefont {Zhang}, \citenamefont
  {Lian}, \citenamefont {Yao}, \citenamefont {Wang}, \citenamefont {Yan},
  \citenamefont {Yi}, \citenamefont {Xu}, \citenamefont {Li}, \citenamefont
  {Hou}, \citenamefont {Xu}, \citenamefont {Guo}, \citenamefont {Zhang},
  \citenamefont {Qi}, \citenamefont {Zhou}, \citenamefont {He},\ and\
  \citenamefont {Duan}}]{guo2024site}%
  \BibitemOpen
  \bibfield  {author} {\bibinfo {author} {\bibfnamefont {S.~A.}\ \bibnamefont
  {Guo}}, \bibinfo {author} {\bibfnamefont {Y.~K.}\ \bibnamefont {Wu}},
  \bibinfo {author} {\bibfnamefont {J.}~\bibnamefont {Ye}}, \bibinfo {author}
  {\bibfnamefont {L.}~\bibnamefont {Zhang}}, \bibinfo {author} {\bibfnamefont
  {W.~Q.}\ \bibnamefont {Lian}}, \bibinfo {author} {\bibfnamefont
  {R.}~\bibnamefont {Yao}}, \bibinfo {author} {\bibfnamefont {Y.}~\bibnamefont
  {Wang}}, \bibinfo {author} {\bibfnamefont {R.~Y.}\ \bibnamefont {Yan}},
  \bibinfo {author} {\bibfnamefont {Y.~J.}\ \bibnamefont {Yi}}, \bibinfo
  {author} {\bibfnamefont {Y.~L.}\ \bibnamefont {Xu}}, \bibinfo {author}
  {\bibfnamefont {B.~W.}\ \bibnamefont {Li}}, \bibinfo {author} {\bibfnamefont
  {Y.~H.}\ \bibnamefont {Hou}}, \bibinfo {author} {\bibfnamefont {Y.~Z.}\
  \bibnamefont {Xu}}, \bibinfo {author} {\bibfnamefont {W.~X.}\ \bibnamefont
  {Guo}}, \bibinfo {author} {\bibfnamefont {C.}~\bibnamefont {Zhang}}, \bibinfo
  {author} {\bibfnamefont {B.~X.}\ \bibnamefont {Qi}}, \bibinfo {author}
  {\bibfnamefont {Z.~C.}\ \bibnamefont {Zhou}}, \bibinfo {author}
  {\bibfnamefont {L.}~\bibnamefont {He}}, \ and\ \bibinfo {author}
  {\bibfnamefont {L.~M.}\ \bibnamefont {Duan}},\ }\href@noop {} {\bibfield
  {journal} {\bibinfo  {journal} {Nature}\ }\textbf {\bibinfo {volume} {630}},\
  \bibinfo {pages} {613} (\bibinfo {year} {2024})}\BibitemShut {NoStop}%
\bibitem [{\citenamefont {Qiao}\ \emph {et~al.}(2024)\citenamefont {Qiao},
  \citenamefont {Cai}, \citenamefont {Wang}, \citenamefont {Du}, \citenamefont
  {Jin}, \citenamefont {Chen}, \citenamefont {Wang}, \citenamefont {Luan},
  \citenamefont {Gao}, \citenamefont {Sun}, \citenamefont {Tian}, \citenamefont
  {Zhang},\ and\ \citenamefont {Kim}}]{qiao2024tunable}%
  \BibitemOpen
  \bibfield  {author} {\bibinfo {author} {\bibfnamefont {M.}~\bibnamefont
  {Qiao}}, \bibinfo {author} {\bibfnamefont {Z.}~\bibnamefont {Cai}}, \bibinfo
  {author} {\bibfnamefont {Y.}~\bibnamefont {Wang}}, \bibinfo {author}
  {\bibfnamefont {B.}~\bibnamefont {Du}}, \bibinfo {author} {\bibfnamefont
  {N.}~\bibnamefont {Jin}}, \bibinfo {author} {\bibfnamefont {W.}~\bibnamefont
  {Chen}}, \bibinfo {author} {\bibfnamefont {P.}~\bibnamefont {Wang}}, \bibinfo
  {author} {\bibfnamefont {C.}~\bibnamefont {Luan}}, \bibinfo {author}
  {\bibfnamefont {E.}~\bibnamefont {Gao}}, \bibinfo {author} {\bibfnamefont
  {X.}~\bibnamefont {Sun}}, \bibinfo {author} {\bibfnamefont {H.}~\bibnamefont
  {Tian}}, \bibinfo {author} {\bibfnamefont {J.}~\bibnamefont {Zhang}}, \ and\
  \bibinfo {author} {\bibfnamefont {K.}~\bibnamefont {Kim}},\ }\href@noop {}
  {\bibfield  {journal} {\bibinfo  {journal} {Nature Physics}\ }\textbf
  {\bibinfo {volume} {20}},\ \bibinfo {pages} {623} (\bibinfo {year}
  {2024})}\BibitemShut {NoStop}%
\bibitem [{\citenamefont {Itano}(2000)}]{itano2000external}%
  \BibitemOpen
  \bibfield  {author} {\bibinfo {author} {\bibfnamefont {W.~M.}\ \bibnamefont
  {Itano}},\ }\href@noop {} {\bibfield  {journal} {\bibinfo  {journal} {Journal
  of Research of the National Institute of Standards and Technology}\ }\textbf
  {\bibinfo {volume} {105}},\ \bibinfo {pages} {829} (\bibinfo {year}
  {2000})}\BibitemShut {NoStop}%
\bibitem [{\citenamefont {Andersson}\ and\ \citenamefont
  {J{\"o}nsson}(2008)}]{andersson2008hfszeeman}%
  \BibitemOpen
  \bibfield  {author} {\bibinfo {author} {\bibfnamefont {M.}~\bibnamefont
  {Andersson}}\ and\ \bibinfo {author} {\bibfnamefont {P.}~\bibnamefont
  {J{\"o}nsson}},\ }\href@noop {} {\bibfield  {journal} {\bibinfo  {journal}
  {Computer Physics Communications}\ }\textbf {\bibinfo {volume} {178}},\
  \bibinfo {pages} {156} (\bibinfo {year} {2008})}\BibitemShut {NoStop}%
\bibitem [{\citenamefont {Cheng}\ and\ \citenamefont
  {Childs}(1985)}]{cheng1985ab}%
  \BibitemOpen
  \bibfield  {author} {\bibinfo {author} {\bibfnamefont {K.}~\bibnamefont
  {Cheng}}\ and\ \bibinfo {author} {\bibfnamefont {W.}~\bibnamefont {Childs}},\
  }\href@noop {} {\bibfield  {journal} {\bibinfo  {journal} {Physical Review
  A}\ }\textbf {\bibinfo {volume} {31}},\ \bibinfo {pages} {2775} (\bibinfo
  {year} {1985})}\BibitemShut {NoStop}%
\bibitem [{\citenamefont {J{\"o}nsson}\ \emph {et~al.}(2022)\citenamefont
  {J{\"o}nsson}, \citenamefont {Godefroid}, \citenamefont {Gaigalas},
  \citenamefont {Ekman}, \citenamefont {Grumer}, \citenamefont {Li},
  \citenamefont {Li}, \citenamefont {Brage}, \citenamefont {Grant},
  \citenamefont {Biero{\'n}} \emph {et~al.}}]{jonsson2022introduction}%
  \BibitemOpen
  \bibfield  {author} {\bibinfo {author} {\bibfnamefont {P.}~\bibnamefont
  {J{\"o}nsson}}, \bibinfo {author} {\bibfnamefont {M.}~\bibnamefont
  {Godefroid}}, \bibinfo {author} {\bibfnamefont {G.}~\bibnamefont {Gaigalas}},
  \bibinfo {author} {\bibfnamefont {J.}~\bibnamefont {Ekman}}, \bibinfo
  {author} {\bibfnamefont {J.}~\bibnamefont {Grumer}}, \bibinfo {author}
  {\bibfnamefont {W.}~\bibnamefont {Li}}, \bibinfo {author} {\bibfnamefont
  {J.}~\bibnamefont {Li}}, \bibinfo {author} {\bibfnamefont {T.}~\bibnamefont
  {Brage}}, \bibinfo {author} {\bibfnamefont {I.~P.}\ \bibnamefont {Grant}},
  \bibinfo {author} {\bibfnamefont {J.}~\bibnamefont {Biero{\'n}}},  \emph
  {et~al.},\ }\href@noop {} {\bibfield  {journal} {\bibinfo  {journal} {Atoms}\
  }\textbf {\bibinfo {volume} {11}},\ \bibinfo {pages} {7} (\bibinfo {year}
  {2022})}\BibitemShut {NoStop}%
\bibitem [{\citenamefont {J{\"o}nsson}\ \emph {et~al.}(2023)\citenamefont
  {J{\"o}nsson}, \citenamefont {Gaigalas}, \citenamefont {Fischer},
  \citenamefont {Biero{\'n}}, \citenamefont {Grant}, \citenamefont {Brage},
  \citenamefont {Ekman}, \citenamefont {Godefroid}, \citenamefont {Grumer},
  \citenamefont {Li} \emph {et~al.}}]{jonsson2023grasp}%
  \BibitemOpen
  \bibfield  {author} {\bibinfo {author} {\bibfnamefont {P.}~\bibnamefont
  {J{\"o}nsson}}, \bibinfo {author} {\bibfnamefont {G.}~\bibnamefont
  {Gaigalas}}, \bibinfo {author} {\bibfnamefont {C.~F.}\ \bibnamefont
  {Fischer}}, \bibinfo {author} {\bibfnamefont {J.}~\bibnamefont {Biero{\'n}}},
  \bibinfo {author} {\bibfnamefont {I.~P.}\ \bibnamefont {Grant}}, \bibinfo
  {author} {\bibfnamefont {T.}~\bibnamefont {Brage}}, \bibinfo {author}
  {\bibfnamefont {J.}~\bibnamefont {Ekman}}, \bibinfo {author} {\bibfnamefont
  {M.}~\bibnamefont {Godefroid}}, \bibinfo {author} {\bibfnamefont
  {J.}~\bibnamefont {Grumer}}, \bibinfo {author} {\bibfnamefont
  {J.}~\bibnamefont {Li}},  \emph {et~al.},\ }\href@noop {} {\bibfield
  {journal} {\bibinfo  {journal} {Atoms}\ }\textbf {\bibinfo {volume} {11}},\
  \bibinfo {pages} {68} (\bibinfo {year} {2023})}\BibitemShut {NoStop}%
\bibitem [{\citenamefont {Parpia}\ and\ \citenamefont
  {Mohanty}(1992)}]{parpia1992relativistic}%
  \BibitemOpen
  \bibfield  {author} {\bibinfo {author} {\bibfnamefont {F.}~\bibnamefont
  {Parpia}}\ and\ \bibinfo {author} {\bibfnamefont {A.}~\bibnamefont
  {Mohanty}},\ }\href@noop {} {\bibfield  {journal} {\bibinfo  {journal}
  {Physical Review A}\ }\textbf {\bibinfo {volume} {46}},\ \bibinfo {pages}
  {3735} (\bibinfo {year} {1992})}\BibitemShut {NoStop}%
\bibitem [{\citenamefont {Grant}(2007)}]{grant2007relativistic}%
  \BibitemOpen
  \bibfield  {author} {\bibinfo {author} {\bibfnamefont {I.~P.}\ \bibnamefont
  {Grant}},\ }\href@noop {} {\bibfield  {journal} {\bibinfo  {journal}
  {Relativistic quantum theory of atoms and molecules: theory and computation}\
  } (\bibinfo {year} {2007})}\BibitemShut {NoStop}%
\bibitem [{\citenamefont {Li}\ \emph {et~al.}(2012)\citenamefont {Li},
  \citenamefont {J{\"o}nsson}, \citenamefont {Godefroid}, \citenamefont
  {Dong},\ and\ \citenamefont {Gaigalas}}]{li2012effects}%
  \BibitemOpen
  \bibfield  {author} {\bibinfo {author} {\bibfnamefont {J.}~\bibnamefont
  {Li}}, \bibinfo {author} {\bibfnamefont {P.}~\bibnamefont {J{\"o}nsson}},
  \bibinfo {author} {\bibfnamefont {M.}~\bibnamefont {Godefroid}}, \bibinfo
  {author} {\bibfnamefont {C.}~\bibnamefont {Dong}}, \ and\ \bibinfo {author}
  {\bibfnamefont {G.}~\bibnamefont {Gaigalas}},\ }\href@noop {} {\bibfield
  {journal} {\bibinfo  {journal} {Physical Review A}\ }\textbf {\bibinfo
  {volume} {86}},\ \bibinfo {pages} {052523} (\bibinfo {year}
  {2012})}\BibitemShut {NoStop}%
\bibitem [{\citenamefont {Zhang}\ \emph {et~al.}(2017)\citenamefont {Zhang},
  \citenamefont {Xie}, \citenamefont {Li},\ and\ \citenamefont
  {Lu}}]{zhang2017theoretical}%
  \BibitemOpen
  \bibfield  {author} {\bibinfo {author} {\bibfnamefont {T.}~\bibnamefont
  {Zhang}}, \bibinfo {author} {\bibfnamefont {L.}~\bibnamefont {Xie}}, \bibinfo
  {author} {\bibfnamefont {J.}~\bibnamefont {Li}}, \ and\ \bibinfo {author}
  {\bibfnamefont {Z.}~\bibnamefont {Lu}},\ }\href@noop {} {\bibfield  {journal}
  {\bibinfo  {journal} {Physical Review A}\ }\textbf {\bibinfo {volume} {96}},\
  \bibinfo {pages} {012514} (\bibinfo {year} {2017})}\BibitemShut {NoStop}%
\bibitem [{\citenamefont {Knecht}\ \emph {et~al.}(2008)\citenamefont {Knecht},
  \citenamefont {Jensen},\ and\ \citenamefont {Fleig}}]{knecht2008large}%
  \BibitemOpen
  \bibfield  {author} {\bibinfo {author} {\bibfnamefont {S.}~\bibnamefont
  {Knecht}}, \bibinfo {author} {\bibfnamefont {H.~J.~A.}\ \bibnamefont
  {Jensen}}, \ and\ \bibinfo {author} {\bibfnamefont {T.}~\bibnamefont
  {Fleig}},\ }\href@noop {} {\bibfield  {journal} {\bibinfo  {journal} {The
  Journal of Chemical Physics}\ }\textbf {\bibinfo {volume} {128}} (\bibinfo
  {year} {2008})}\BibitemShut {NoStop}%
\bibitem [{\citenamefont {Jensen}\ \emph {et~al.}(2022)\citenamefont {Jensen},
  \citenamefont {Bast}, \citenamefont {Gomes}, \citenamefont {Saue},
  \citenamefont {Visscher}, \citenamefont {Aucar}, \citenamefont {Bakken},
  \citenamefont {Chibueze}, \citenamefont {Creutzberg}, \citenamefont {Dyall},
  \citenamefont {Dubillard}, \citenamefont {Ekström}, \citenamefont {Eliav},
  \citenamefont {Enevoldsen}, \citenamefont {Faßhauer}, \citenamefont {Fleig},
  \citenamefont {Fossgaard}, \citenamefont {Halbert}, \citenamefont
  {Hedegård}, \citenamefont {Helgaker}, \citenamefont {Helmich-Paris},
  \citenamefont {Henriksson}, \citenamefont {van Horn}, \citenamefont {Iliaš},
  \citenamefont {Jacob}, \citenamefont {Knecht}, \citenamefont {Komorovský},
  \citenamefont {Kullie}, \citenamefont {Lærdahl}, \citenamefont {Larsen},
  \citenamefont {Lee}, \citenamefont {List}, \citenamefont {Nataraj},
  \citenamefont {Nayak}, \citenamefont {Norman}, \citenamefont {Olejniczak},
  \citenamefont {Olsen}, \citenamefont {Olsen}, \citenamefont {Papadopoulos},
  \citenamefont {Park}, \citenamefont {Pedersen}, \citenamefont {Pernpointner},
  \citenamefont {Pototschnig}, \citenamefont {Di~Remigio}, \citenamefont
  {Repiský}, \citenamefont {Ruud}, \citenamefont {Sałek}, \citenamefont
  {Schimmelpfennig}, \citenamefont {Senjean}, \citenamefont {Shee},
  \citenamefont {Sikkema}, \citenamefont {Sunaga}, \citenamefont {Thorvaldsen},
  \citenamefont {Thyssen}, \citenamefont {van Stralen}, \citenamefont {Vidal},
  \citenamefont {Villaume}, \citenamefont {Visser}, \citenamefont {Winther},
  \citenamefont {Yamamoto},\ and\ \citenamefont
  {Yuan}}]{jensen_h_j_aa_2022_6010450}%
  \BibitemOpen
  \bibfield  {author} {\bibinfo {author} {\bibfnamefont {H.~J.~A.}\
  \bibnamefont {Jensen}}, \bibinfo {author} {\bibfnamefont {R.}~\bibnamefont
  {Bast}}, \bibinfo {author} {\bibfnamefont {A.~S.~P.}\ \bibnamefont {Gomes}},
  \bibinfo {author} {\bibfnamefont {T.}~\bibnamefont {Saue}}, \bibinfo {author}
  {\bibfnamefont {L.}~\bibnamefont {Visscher}}, \bibinfo {author}
  {\bibfnamefont {I.~A.}\ \bibnamefont {Aucar}}, \bibinfo {author}
  {\bibfnamefont {V.}~\bibnamefont {Bakken}}, \bibinfo {author} {\bibfnamefont
  {C.}~\bibnamefont {Chibueze}}, \bibinfo {author} {\bibfnamefont
  {J.}~\bibnamefont {Creutzberg}}, \bibinfo {author} {\bibfnamefont {K.~G.}\
  \bibnamefont {Dyall}}, \bibinfo {author} {\bibfnamefont {S.}~\bibnamefont
  {Dubillard}}, \bibinfo {author} {\bibfnamefont {U.}~\bibnamefont {Ekström}},
  \bibinfo {author} {\bibfnamefont {E.}~\bibnamefont {Eliav}}, \bibinfo
  {author} {\bibfnamefont {T.}~\bibnamefont {Enevoldsen}}, \bibinfo {author}
  {\bibfnamefont {E.}~\bibnamefont {Faßhauer}}, \bibinfo {author}
  {\bibfnamefont {T.}~\bibnamefont {Fleig}}, \bibinfo {author} {\bibfnamefont
  {O.}~\bibnamefont {Fossgaard}}, \bibinfo {author} {\bibfnamefont
  {L.}~\bibnamefont {Halbert}}, \bibinfo {author} {\bibfnamefont {E.~D.}\
  \bibnamefont {Hedegård}}, \bibinfo {author} {\bibfnamefont {T.}~\bibnamefont
  {Helgaker}}, \bibinfo {author} {\bibfnamefont {B.}~\bibnamefont
  {Helmich-Paris}}, \bibinfo {author} {\bibfnamefont {J.}~\bibnamefont
  {Henriksson}}, \bibinfo {author} {\bibfnamefont {M.}~\bibnamefont {van
  Horn}}, \bibinfo {author} {\bibfnamefont {M.}~\bibnamefont {Iliaš}},
  \bibinfo {author} {\bibfnamefont {C.~R.}\ \bibnamefont {Jacob}}, \bibinfo
  {author} {\bibfnamefont {S.}~\bibnamefont {Knecht}}, \bibinfo {author}
  {\bibfnamefont {S.}~\bibnamefont {Komorovský}}, \bibinfo {author}
  {\bibfnamefont {O.}~\bibnamefont {Kullie}}, \bibinfo {author} {\bibfnamefont
  {J.~K.}\ \bibnamefont {Lærdahl}}, \bibinfo {author} {\bibfnamefont {C.~V.}\
  \bibnamefont {Larsen}}, \bibinfo {author} {\bibfnamefont {Y.~S.}\
  \bibnamefont {Lee}}, \bibinfo {author} {\bibfnamefont {N.~H.}\ \bibnamefont
  {List}}, \bibinfo {author} {\bibfnamefont {H.~S.}\ \bibnamefont {Nataraj}},
  \bibinfo {author} {\bibfnamefont {M.~K.}\ \bibnamefont {Nayak}}, \bibinfo
  {author} {\bibfnamefont {P.}~\bibnamefont {Norman}}, \bibinfo {author}
  {\bibfnamefont {G.}~\bibnamefont {Olejniczak}}, \bibinfo {author}
  {\bibfnamefont {J.}~\bibnamefont {Olsen}}, \bibinfo {author} {\bibfnamefont
  {J.~M.~H.}\ \bibnamefont {Olsen}}, \bibinfo {author} {\bibfnamefont
  {A.}~\bibnamefont {Papadopoulos}}, \bibinfo {author} {\bibfnamefont {Y.~C.}\
  \bibnamefont {Park}}, \bibinfo {author} {\bibfnamefont {J.~K.}\ \bibnamefont
  {Pedersen}}, \bibinfo {author} {\bibfnamefont {M.}~\bibnamefont
  {Pernpointner}}, \bibinfo {author} {\bibfnamefont {J.~V.}\ \bibnamefont
  {Pototschnig}}, \bibinfo {author} {\bibfnamefont {R.}~\bibnamefont
  {Di~Remigio}}, \bibinfo {author} {\bibfnamefont {M.}~\bibnamefont
  {Repiský}}, \bibinfo {author} {\bibfnamefont {K.}~\bibnamefont {Ruud}},
  \bibinfo {author} {\bibfnamefont {P.}~\bibnamefont {Sałek}}, \bibinfo
  {author} {\bibfnamefont {B.}~\bibnamefont {Schimmelpfennig}}, \bibinfo
  {author} {\bibfnamefont {B.}~\bibnamefont {Senjean}}, \bibinfo {author}
  {\bibfnamefont {A.}~\bibnamefont {Shee}}, \bibinfo {author} {\bibfnamefont
  {J.}~\bibnamefont {Sikkema}}, \bibinfo {author} {\bibfnamefont
  {A.}~\bibnamefont {Sunaga}}, \bibinfo {author} {\bibfnamefont {A.~J.}\
  \bibnamefont {Thorvaldsen}}, \bibinfo {author} {\bibfnamefont
  {J.}~\bibnamefont {Thyssen}}, \bibinfo {author} {\bibfnamefont
  {J.}~\bibnamefont {van Stralen}}, \bibinfo {author} {\bibfnamefont {M.~L.}\
  \bibnamefont {Vidal}}, \bibinfo {author} {\bibfnamefont {S.}~\bibnamefont
  {Villaume}}, \bibinfo {author} {\bibfnamefont {O.}~\bibnamefont {Visser}},
  \bibinfo {author} {\bibfnamefont {T.}~\bibnamefont {Winther}}, \bibinfo
  {author} {\bibfnamefont {S.}~\bibnamefont {Yamamoto}}, \ and\ \bibinfo
  {author} {\bibfnamefont {X.}~\bibnamefont {Yuan}},\ }\href {\doibase
  10.5281/zenodo.6010450} {\  (\bibinfo {year} {2022}),\
  10.5281/zenodo.6010450},\ \bibinfo {note} {{Please join our mailing list:
  https://groups.google.com/g/dirac-users}}\BibitemShut {NoStop}%
\bibitem [{\citenamefont {Gomes}\ \emph {et~al.}(2010)\citenamefont {Gomes},
  \citenamefont {Dyall},\ and\ \citenamefont
  {Visscher}}]{gomes2010relativistic}%
  \BibitemOpen
  \bibfield  {author} {\bibinfo {author} {\bibfnamefont {A.~S.}\ \bibnamefont
  {Gomes}}, \bibinfo {author} {\bibfnamefont {K.~G.}\ \bibnamefont {Dyall}}, \
  and\ \bibinfo {author} {\bibfnamefont {L.}~\bibnamefont {Visscher}},\
  }\href@noop {} {\bibfield  {journal} {\bibinfo  {journal} {Theoretical
  Chemistry Accounts}\ }\textbf {\bibinfo {volume} {127}},\ \bibinfo {pages}
  {369} (\bibinfo {year} {2010})}\BibitemShut {NoStop}%
\bibitem [{\citenamefont {Hubert}\ and\ \citenamefont
  {Fleig}(2022)}]{hubert2022electric}%
  \BibitemOpen
  \bibfield  {author} {\bibinfo {author} {\bibfnamefont {M.}~\bibnamefont
  {Hubert}}\ and\ \bibinfo {author} {\bibfnamefont {T.}~\bibnamefont {Fleig}},\
  }\href@noop {} {\bibfield  {journal} {\bibinfo  {journal} {Physical Review
  A}\ }\textbf {\bibinfo {volume} {106}},\ \bibinfo {pages} {022817} (\bibinfo
  {year} {2022})}\BibitemShut {NoStop}%
\bibitem [{\citenamefont {Fleig}\ \emph {et~al.}(2001)\citenamefont {Fleig},
  \citenamefont {Olsen},\ and\ \citenamefont {Marian}}]{fleig2001generalized}%
  \BibitemOpen
  \bibfield  {author} {\bibinfo {author} {\bibfnamefont {T.}~\bibnamefont
  {Fleig}}, \bibinfo {author} {\bibfnamefont {J.}~\bibnamefont {Olsen}}, \ and\
  \bibinfo {author} {\bibfnamefont {C.~M.}\ \bibnamefont {Marian}},\
  }\href@noop {} {\bibfield  {journal} {\bibinfo  {journal} {The Journal of
  Chemical Physics}\ }\textbf {\bibinfo {volume} {114}},\ \bibinfo {pages}
  {4775} (\bibinfo {year} {2001})}\BibitemShut {NoStop}%
\bibitem [{\citenamefont {Fleig}\ \emph {et~al.}(2003)\citenamefont {Fleig},
  \citenamefont {Olsen},\ and\ \citenamefont
  {Visscher}}]{fleig2003generalized}%
  \BibitemOpen
  \bibfield  {author} {\bibinfo {author} {\bibfnamefont {T.}~\bibnamefont
  {Fleig}}, \bibinfo {author} {\bibfnamefont {J.}~\bibnamefont {Olsen}}, \ and\
  \bibinfo {author} {\bibfnamefont {L.}~\bibnamefont {Visscher}},\ }\href@noop
  {} {\bibfield  {journal} {\bibinfo  {journal} {The Journal of Chemical
  Physics}\ }\textbf {\bibinfo {volume} {119}},\ \bibinfo {pages} {2963}
  (\bibinfo {year} {2003})}\BibitemShut {NoStop}%
\bibitem [{\citenamefont {Fleig}\ \emph {et~al.}(2006)\citenamefont {Fleig},
  \citenamefont {Jensen}, \citenamefont {Olsen},\ and\ \citenamefont
  {Visscher}}]{fleig2006generalized}%
  \BibitemOpen
  \bibfield  {author} {\bibinfo {author} {\bibfnamefont {T.}~\bibnamefont
  {Fleig}}, \bibinfo {author} {\bibfnamefont {H.~J.~A.}\ \bibnamefont
  {Jensen}}, \bibinfo {author} {\bibfnamefont {J.}~\bibnamefont {Olsen}}, \
  and\ \bibinfo {author} {\bibfnamefont {L.}~\bibnamefont {Visscher}},\
  }\href@noop {} {\bibfield  {journal} {\bibinfo  {journal} {The Journal of
  Chemical Physics}\ }\textbf {\bibinfo {volume} {124}} (\bibinfo {year}
  {2006})}\BibitemShut {NoStop}%
\bibitem [{\citenamefont {Vanier}\ and\ \citenamefont
  {Audoin}(1989)}]{vanier1989quantum}%
  \BibitemOpen
  \bibfield  {author} {\bibinfo {author} {\bibfnamefont {J.}~\bibnamefont
  {Vanier}}\ and\ \bibinfo {author} {\bibfnamefont {C.}~\bibnamefont
  {Audoin}},\ }\href@noop {} {\bibfield  {journal} {\bibinfo  {journal} {The
  quantum physics of atomic frequency standards}\ } (\bibinfo {year}
  {1989})}\BibitemShut {NoStop}%
\bibitem [{\citenamefont {Safronova}\ \emph {et~al.}(2011)\citenamefont
  {Safronova}, \citenamefont {Kozlov},\ and\ \citenamefont
  {Clark}}]{safronova2011precision}%
  \BibitemOpen
  \bibfield  {author} {\bibinfo {author} {\bibfnamefont {M.}~\bibnamefont
  {Safronova}}, \bibinfo {author} {\bibfnamefont {M.}~\bibnamefont {Kozlov}}, \
  and\ \bibinfo {author} {\bibfnamefont {C.~W.}\ \bibnamefont {Clark}},\
  }\href@noop {} {\bibfield  {journal} {\bibinfo  {journal} {Physical Review
  Letters}\ }\textbf {\bibinfo {volume} {107}},\ \bibinfo {pages} {143006}
  (\bibinfo {year} {2011})}\BibitemShut {NoStop}%
\bibitem [{\citenamefont {Bohman}\ \emph {et~al.}(2023)\citenamefont {Bohman},
  \citenamefont {Porsev}, \citenamefont {Hume}, \citenamefont {Leibrandt},\
  and\ \citenamefont {Safronova}}]{bohman2023enhancing}%
  \BibitemOpen
  \bibfield  {author} {\bibinfo {author} {\bibfnamefont {M.~A.}\ \bibnamefont
  {Bohman}}, \bibinfo {author} {\bibfnamefont {S.~G.}\ \bibnamefont {Porsev}},
  \bibinfo {author} {\bibfnamefont {D.~B.}\ \bibnamefont {Hume}}, \bibinfo
  {author} {\bibfnamefont {D.~R.}\ \bibnamefont {Leibrandt}}, \ and\ \bibinfo
  {author} {\bibfnamefont {M.~S.}\ \bibnamefont {Safronova}},\ }\href@noop {}
  {\bibfield  {journal} {\bibinfo  {journal} {Physical Review A}\ }\textbf
  {\bibinfo {volume} {108}},\ \bibinfo {pages} {053120} (\bibinfo {year}
  {2023})}\BibitemShut {NoStop}%
\bibitem [{\citenamefont {Tang}\ \emph {et~al.}(2024)\citenamefont {Tang},
  \citenamefont {Wei}, \citenamefont {Sahoo}, \citenamefont {Li}, \citenamefont
  {Yang}, \citenamefont {Zou},\ and\ \citenamefont
  {Huang}}]{tang2024blackbody}%
  \BibitemOpen
  \bibfield  {author} {\bibinfo {author} {\bibfnamefont {Z.-M.}\ \bibnamefont
  {Tang}}, \bibinfo {author} {\bibfnamefont {Y.-F.}\ \bibnamefont {Wei}},
  \bibinfo {author} {\bibfnamefont {B.}~\bibnamefont {Sahoo}}, \bibinfo
  {author} {\bibfnamefont {C.-B.}\ \bibnamefont {Li}}, \bibinfo {author}
  {\bibfnamefont {Y.}~\bibnamefont {Yang}}, \bibinfo {author} {\bibfnamefont
  {Y.}~\bibnamefont {Zou}}, \ and\ \bibinfo {author} {\bibfnamefont {X.-R.}\
  \bibnamefont {Huang}},\ }\href@noop {} {\bibfield  {journal} {\bibinfo
  {journal} {Physical Review A}\ }\textbf {\bibinfo {volume} {110}},\ \bibinfo
  {pages} {043108} (\bibinfo {year} {2024})}\BibitemShut {NoStop}%
\bibitem [{\citenamefont {Sahoo}\ and\ \citenamefont
  {Kumar}(2017)}]{sahoo2017relativistic}%
  \BibitemOpen
  \bibfield  {author} {\bibinfo {author} {\bibfnamefont {B.}~\bibnamefont
  {Sahoo}}\ and\ \bibinfo {author} {\bibfnamefont {P.}~\bibnamefont {Kumar}},\
  }\href@noop {} {\bibfield  {journal} {\bibinfo  {journal} {Physical Review
  A}\ }\textbf {\bibinfo {volume} {96}},\ \bibinfo {pages} {012511} (\bibinfo
  {year} {2017})}\BibitemShut {NoStop}%
\bibitem [{\citenamefont {Nandy}\ and\ \citenamefont
  {Sahoo}(2014)}]{nandy2014quadrupole}%
  \BibitemOpen
  \bibfield  {author} {\bibinfo {author} {\bibfnamefont {D.~K.}\ \bibnamefont
  {Nandy}}\ and\ \bibinfo {author} {\bibfnamefont {B.}~\bibnamefont {Sahoo}},\
  }\href@noop {} {\bibfield  {journal} {\bibinfo  {journal} {Physical Review
  A}\ }\textbf {\bibinfo {volume} {90}},\ \bibinfo {pages} {050503} (\bibinfo
  {year} {2014})}\BibitemShut {NoStop}%
\bibitem [{\citenamefont {Stone}(2016)}]{stone2016table}%
  \BibitemOpen
  \bibfield  {author} {\bibinfo {author} {\bibfnamefont {N.}~\bibnamefont
  {Stone}},\ }\href@noop {} {\bibfield  {journal} {\bibinfo  {journal} {Atomic
  Data and Nuclear Data Tables}\ }\textbf {\bibinfo {volume} {111}},\ \bibinfo
  {pages} {1} (\bibinfo {year} {2016})}\BibitemShut {NoStop}%
\end{thebibliography}%

\end{document}